\documentclass[fleqn,usenatbib]{mnras}
\usepackage{comment}
\usepackage{newtxtext,newtxmath}

\usepackage[T1]{fontenc}
\usepackage{ae,aecompl}

\newcommand{\kps}{km s$^{-1}$} 
\newcommand{\mps}{m s$^{-1}$}
\newcommand{\tess}{\textit{TESS}}
\newcommand{\me}{$M_{\oplus}$}
\newcommand{\re}{$R_{\oplus}$}
\newcommand{\gcm}{g cm$^{-3}$}
\newcommand{\msun}{$M_{\odot}$}

\newcommand{\planet}{TOI-132\,b }
\newcommand{\ms}{\ensuremath{m\,s^{-1}}}
\newcommand{\Gaia}{{\it Gaia}}
\newcommand{\cms}{cm s$^{-2}$}
\newcommand{\Kepler}{\textit{Kepler }}
\newcommand{\juliet}{\texttt{juliet}}

\usepackage{graphicx}	
\usepackage{amsmath}	
\usepackage{amssymb}	

\title[A short-period Neptune transiting TOI-132]{TOI-132\,b: A short-period planet in the Neptune desert transiting a $V=11.3$ G-type star}

\author[M.~R. D\'iaz et al.]{
Mat\'ias R. D\'iaz,$^{1}$\thanks{E-mail: matias.diaz.m@ug.uchile.cl}
James S. Jenkins,$^{1}$
Davide Gandolfi,$^{2}$ 
Eric D. Lopez,$^{3}$
\newauthor
~Maritza G. Soto, $^{4}$
P\'ia Cort\'es-Zuleta,$^{1}$
Zaira M. Berdi\~{n}as,$^{1}$
Keivan G. Stassun,$^{5,6}$
\newauthor
~Karen A. Collins,$^{7}$
Jos\'e I. Vines,$^{1}$
Carl Ziegler,$^{8}$
Malcom Fridlund,$^{9,10}$
Eric L.N. Jensen,$^{11}$  
\newauthor
~Felipe Murgas,$^{12,13}$
Alexandre Santerne,$^{14}$
Paul A. Wilson,$^{15}$
Massimiliano Esposito,$^{16}$
\newauthor
~Artie P. Hatzes,$^{16}$
Marshall C. Johnson,$^{17}$
Kristine W. F. Lam,$^{18}$
John H. Livingston,$^{19}$
\newauthor 
~Vincent Van Eylen,$^{20,21}$
Norio Narita,$^{12,22,23,24}$
Cesar Brice\~{n}o,$^{25}$
Kevin I.\ Collins,$^{26}$
\newauthor
~Szilard Csizmadia,$^{27}$
Michael Fausnaugh,$^{28}$
Tianjun Gan,$^{29}$ 
Iska Georgieva,$^{9}$
\newauthor
~Ana Glidden,$^{28,30}$
Jon M. Jenkins,$^{31}$
David W. Latham,$^{7}$
Nicholas M. Law,$^{32}$
\newauthor
~Andrew W. Mann,$^{32}$
Savita Mathur,$^{12}$
Ismael Mireles,$^{28}$
Robert Morris,$^{31,33}$
\newauthor
~Enric Pall\'{e},$^{12,13}$
Carina M. Persson,$^{9}$
Stephen Rinehart,$^{3}$
Mark E. Rose,$^{31}$
\newauthor
~Sara Seager,$^{28,30,34}$
Jeffrey C. Smith,$^{31,33}$
Thiam-Guan Tan,$^{35}$
Andrei Tokovinin,$^{25}$
\newauthor
~Andrew Vanderburg,$^{36}$
Roland Vanderspek,$^{28}$
Daniel A. Yahalomi$^{7}$
\\
Authors' affiliations are shown at the end of the manuscript}

\date{Accepted XXX. Received YYY; in original form YYZ}
\pubyear{2019}
\begin{document}
\label{firstpage}
\pagerange{\pageref{firstpage}--\pageref{lastpage}}
\maketitle

\begin{abstract}
The Neptune desert is a feature seen in the radius-mass-period plane, whereby a notable dearth of short period, Neptune-like planets is found. Here we report the \tess\, discovery of a new short-period planet in the Neptune desert, orbiting the G-type dwarf TYC\,8003-1117-1 (TOI-132).  \tess\, photometry shows transit-like dips at the level of $\sim$1400\,ppm occurring every $\sim$2.11 days. High-precision radial velocity follow-up with HARPS confirmed the planetary nature of the transit signal and provided a semi-amplitude radial velocity variation of $\sim$11.5~\ms, which, when combined with the stellar mass of $0.97\pm0.06$~\msun, provides a planetary mass of 22.83$^{+1.81}_{-1.80}$~\me. Modeling the \tess\, high-quality light curve returns a planet radius of 3.43$^{+0.13}_{-0.14}$~\re, and therefore the planet bulk density is found to be 3.11$^{+0.44}_{-0.450}$~\gcm. Planet structure models suggest that the bulk of the planet mass is in the form of a rocky core, with an atmospheric mass fraction of 4.3$^{+1.2}_{-2.3}$\%.  \planet\, is a \tess\, Level 1 Science Requirement candidate, and therefore priority follow-up will allow the search for additional planets in the system, whilst helping to constrain low-mass planet formation and evolution models, particularly valuable for better understanding the Neptune desert.
\end{abstract}

\section{Introduction}
The \Kepler space telescope \citep{Borucki2010} has allowed us to understand the  population of small planets ($R_{\rm p}\,<\,4$~\re) in a real statistical sense for the first time. \Kepler revealed that the majority of planets are the so-called super-Earths, with an occurrence rate of $\sim$6\% of Earth-size planets around Sun-like stars \citep{Petigura2013}. \Kepler has also unveiled a bimodality in the radius distribution of such planets \citep{Fulton2017, VanEylen2018}, which could be the result of photo-evaporation of the planetary atmosphere due to the intense stellar radiation \citep{LopezFortney2013,OwenWu2013,Jin2014,ChenRogers2016}. Furthermore, higher-mass planets in the Neptune regime are also more abundant than the large gas giant planets. It is important to note that the distinction between super-Earth and sub-Neptune is based only on the radius, where the first class is commonly defined as planets with $1\,R_{\oplus}\,<\,R_{\rm p}\,<\,2\,R_{\oplus}$ while the latter comprises planets with $2 \,R_{\oplus}\,<\,R_{\rm p}\,<\,4\,R_{\oplus}$. From \Kepler statistics, 25\% of Sun-like stars in our galaxy are found to host at least one small planet ($R_{\rm p}<4R_{\oplus}$) on a short period orbit ($P<100$\,d) \citep{Batalha2013, Marcy2014}.

Although Neptune-sized planets orbiting Sun-like stars are fairly abundant \citep[e.g.,][]{Espinoza2016,Luque2019,Mayo2019,Palle2019}, at short orbital periods they are very rare. A number of early studies indicated a lack of Neptune-sized planets with periods shorter than 2$\--$4~days \citep{SzaboKiss2011, BenitezLlambay2011, BeaugeNesvorny2013, Helled2016}, and the term ``Neptune desert" was coined to explain this paucity. \citet{Mazeh2016} placed this dearth on a statistical footing, whilst providing robust boundaries for the region. Even though the dominant mechanism that produces this desert is currently unknown, models that invoke tidal disruption of a high-eccentricity migration planet, coupled with photoevaporation can explain the triangular shape of the gap described by \cite{Mazeh2016} (see also \citealt{Lundkvist2016,OwenLai2018}). 

The Neptune desert may be a region of parameter space with a paucity of such planets, but it is not completely empty. \citet{West2019} discovered the planet NGTS-4$b$ as part of the Next Generation Transit Survey \citep{Wheatley2018}. Although the star is fairly faint ($V=13.14$), making the constraints on the radius and mass difficult, the planet resides inside the boundaries of the desert, as defined by \citet{Mazeh2016}. A more recent example was found using data from the Transiting Exoplanet Survey Satellite \citep[{\it TESS};][]{Ricker2015}: a planet orbiting the star HD\,21966 residing in the edge of this region \citep{Esposito2019}. The primary goal of \tess\, is to discover 50 planets with radii $\le$\,4~\re\ transiting stars brighter than V\,$\le$\,12, for which precise masses can be measured using high-precision Doppler spectroscopy, better constraining the planetary bulk density. In doing so, the mission is also providing unprecedented targets to follow-up to study the Neptune desert, particularly the discovery of the first ultra hot Neptune, LTT\,9779\,b (Jenkins et al. 2019).  This planet resides on the edge of the Neptune desert, and since the star is bright ($V=9.76$), detailed follow-up can be performed to shed light on the processes that sculpt the desert. However, more such examples are necessary in order to uncover the dominant process(es) at play.

Here we present the discovery of TOI-132\,b, a 22.8-\me\ Neptune-sized planet discovered by  \tess\, and confirmed using high-precision Doppler spectroscopy from the High Accuracy Radial velocity Planet Searcher \citep[HARPS;][]{Pepe2002}. 

\section{Photometry}
\subsection{\tess\, Photometry}
TYC\,8003-1117-1 (also known as TOI-132) was observed by \tess\, in Sector 1 on Camera 2 in short-cadence mode ($T_\mathrm{exp}$\,=\,2 minutes).  The total time baseline of the observations is 27.87\,days, spanning from July 25th to August 22nd 2018. TOI-132\,b was identified as a potential transiting planet signature by the Science Processing Operations Center (SPOC) in the transit search run on Sector 1 \citep{JonJenkins2002, JonJenkins2010} and promoted to TOI status by the TESS Science Office based on the SPOC Data Validation (DV) reports \citep{Twicken2018, Li2019}. 

The target was selected from the \tess\, alerts website\footnote{\url{//https://tev.mit.edu/data/}}, based on the magnitude of the star ($V$=11.2 mag) and period of the candidate, since it presented a good opportunity to be confirmed relatively quickly with HARPS.

\begin{figure*}
\centering
\includegraphics[scale=1]{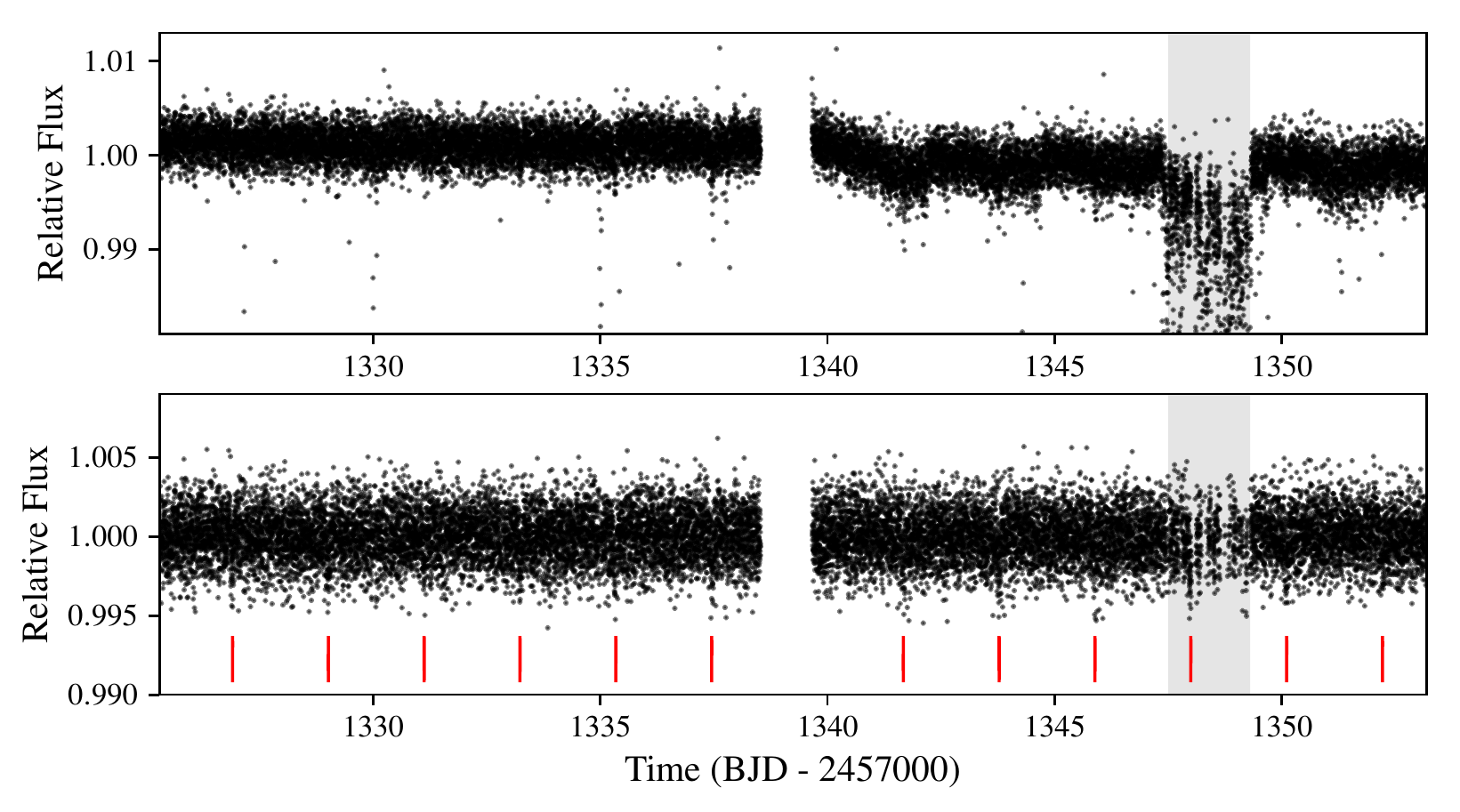}
\caption{\tess~ light curve for TOI-132. Top panel shows the Simple Aperture Photometry (\texttt{SAP\_FLUX}). Bottom panel shows the systematic-corrected \texttt{PDCSAP\_FLUX} photometry after normalizing by the median and rejecting 4$\sigma$ values. Red vertical lines show the position of the 12 transits identified in the TESS alert from Sector 1.}\label{fig:tesslc}
\end{figure*}

In addition, the DV report for TOI-132\,b is very clean, and the planetary signature passed all of the diagnostic tests conducted by DV, including the odd/even depth test, the weak secondary test, the ghost diagnostic test, the difference image centroid shift test.
We retrieved the photometry provided by the \tess~ SPOC pipeline \citep{JonJenkins2016}, and accessed the data from the simple aperture photometry (\texttt{SAP\_FLUX}) and the Presearch Data Conditioning simple aperture photometry (\texttt{PDCSAP\_FLUX}, \citealt{Smith2012,Stumpe2014}), which contains systematics-corrected data using the algorithms previously used for {\it Kepler} \citep{JonJenkins2017}. The median-normalized \texttt{SAP\_FLUX} photometry is shown in the top panel of Figure~\ref{fig:tesslc}. Bottom panel shows the \texttt{PDCSAP\_FLUX} photometry, divided by its median value and after applying a 4-$\sigma$ clipping rejection. This light curve is used throughout all the analyses in this paper. The gap in the middle of the time series occurred when the observations were stopped to allow for the data down-link. Finally, in order to avoid any bias in our analysis, we excluded the photometric measurements between (BJD - 2457000) 1347.5 and 1349.3 (gray shaded area) given that the spacecraft pointing jitter was higher than nominal, as described by \citet{Huang2018} and also noted in recent \tess~ discoveries (see, e.g., \citealt[][]{Espinoza2019a}). A total of 11 transit events were considered for further analysis in the present work. Magnitudes and stellar parameters for TOI-132 are shown in Table~\ref{tab:star} (see also Section \ref{sec:star}).

\begin{table*}
    \centering
    \caption{Stellar parameters for TOI-132.}\label{tab:star}
    \begin{tabular}{ccc}
    \hline\hline
    Parameter     & Value & Source \\
     \hline \hline
    \tess~ Names & TIC89020549 (TOI-132.01) &  \\
    RA (hh:mm:ss) & 22:33:35.8683 & \Gaia \\
    Dec (dd:mm:ss) & -43:26:11.9167& \Gaia \\
    $\mu$ RA (mas yr$^{-1}$) & 35.553 $\pm$ 0.043 &\Gaia \\
    $\mu$  (mas yr$^{-1}$)& -53.055 $\pm$ 0.054 & \Gaia\\
    Parallax (mas) & 6.08 $\pm$ 0.04 & \Gaia*\\
    Distance (pc) & 164.47 $\pm$  27.32 & \Gaia \\
    SpT & G8V & This work \\
    \hline
    Photometry & &\\
     $B$ & 12.07 $\pm$0.17 &Tycho-2\\
     $V$ & 11.29 $\pm$ 0.07 & Tycho-2 \\
     $g$&11.85 $\pm$ 0.02 &APASS\\
     $r$&11.24 $\pm$ 0.01 &APASS\\
     $i$& 11.08 $\pm$ 0.02 &APASS\\
     {\it TESS} &10.80 $\pm$ 0.02 & \citet{Stassun2018b}\\
      $Gaia$ & 11.2935 $\pm$ 0.0003 & \Gaia \\
      $J$  & 10.14 $\pm$0.02 &2MASS\\
      $H$ & 9.76 $\pm$ 0.02&2MASS\\
      $K_{s}$ & 9.65 $\pm$ 0.02 &2MASS\\
      $W_{1}$& 9.61$\pm$ 0.02&WISE \\
      $W_{2}$& 9.69$\pm$ 0.02&WISE \\
      $W_{3}$& 9.60$\pm$0.04 &WISE \\
      $W_{4}$& 8.72$\pm$0.42 &WISE \\
\hline
    Derived Parameters&&\\
    $T_{\rm eff}$ (K)    &5397 $ \pm$ 46  & This work\\
    log\,$g$ (\cms)    & $4.48 \pm 0.23$  & This work\\
    $\left[\rm{Fe/H}\right]$	 (dex)    & $0.16 \pm 0.10$     &This work\\
    $L (L_{\odot}$) & 0.60 $\pm$ 0.05 & This work \\
    $R~(R_{\odot}$) & 0.90 $\pm$ 0.02 & This work\\
    $M~(M_{\odot}$) & 0.97 $\pm$ 0.06 & This work\\
    $v~sin(i)$ (\kps) & 3.00 $\pm$ 0.30 & This work\\
    $v_{\rm mac}$ (\kps) & 1.74 $\pm$ 0.20 & This work\\
    $\rho_{\star}$ (g cm$^{-3}$) & 1.89 $\pm$ 0.15 & This work\\
    $\log R_{HK}$ (dex) & $-$5.02 $\pm$ 0.13 & This work\\ 
    Age (Gyr) & 6.34$^{+0.42}_{-2.35}$& This work\\ 
    (U,V,W) (km s$^{-1}$)&18.4 $\pm$ 0.2, -32.6 $\pm$ 0.4, 16.5 $\pm$ 0.4& This work\\ \hline
         \hline
    \multicolumn{2}{l}{*Correction of +82 $\mu$as from \citet{StassunTorres2018} applied to the \Gaia\, value.}
    \end{tabular}
\end{table*}

\subsection{Ground-based time-series photometry}

We acquired ground-based time-series follow-up photometry of TOI-132 as part of the \tess\, Follow-up Observing Program (TFOP) to attempt to rule out nearby eclipsing binaries (NEBs) in all stars that could be blended in the \tess\, aperture as potential sources of the \tess\, detection. Furthermore, we attempt to $i)$ detect the transit-like event on target to confirm the event depth and thus the \tess\ photometric deblending factor, $ii)$ refine the \tess\, ephemeris, $iii)$ provide additional epochs of transit center time measurements to supplement the transit timing variation (TTV) analysis, and $iv)$ place constraints on transit depth differences across filter bands. We used the {\tt TESS Transit Finder}, which is a customized version of the {\tt Tapir} software package \citep{Jensen2013}, to schedule our transit observations. 

We observed TOI-132 continuously for 443 minutes on UTC 2018 September 09 in $\rm R_c$ band from the Perth Exoplanet Survey Telescope (PEST) near Perth, Australia. The 0.3-m telescope is equipped with a $1530\times1020$ pixels SBIG ST-8XME camera with an image scale of 1$\farcs$2 pixel$^{-1}$ resulting in a $31\arcmin\times21\arcmin$ field of view. A custom pipeline was used to calibrate the images and extract the differential photometry using an aperture with radius 8$\farcs$2. The images have typical stellar point spread functions (PSFs) with a full width at half maximum (FWHM) of $\sim$4$\arcsec$.  The data rule out NEBs in stars within $2\farcm5$ of the target star that are fainter by as much as 6.4 magnitudes in $\rm R_c$ band. 

We also observed full predicted transit durations of TOI-132 continuously in z-short band on UTC 2018 November 14, UTC 2019 June 19, and UTC 2019 July 06 from the Las Cumbres Observatory (LCO) 1.0 m telescopes \citep{Brown2013} at Cerro Tololo Inter-American Observatory for 277, 335, and 283 minutes, respectively. Another full transit was observed continuously for 232 minutes in B-band on UTC 2019 August 02 from an LCO 1.0 m telescope at Siding Spring Observatory. The $4096\times4096$ LCO SINISTRO cameras have an image scale of 0$\farcs$389 pixel$^{-1}$ resulting in a $26\arcmin\times26\arcmin$ field of view. The images were calibrated by the standard LCO BANZAI pipeline \citep{McCully2018} and the photometric data were extracted using the {\tt AstroImageJ} ({\tt AIJ}) software package \citep{Collins2017}.

The November data rule out NEBs in all stars within $2\farcm5$ of the target star that are fainter by as much as 8.7 magnitudes in z-short band, which includes all known \Gaia\ DR2 stars that are blended in the \tess\, aperture. The June observation confirmed a $\sim$1400 ppm deep ingress on target arriving $\sim$80 minutes late relative to the original TOI ephemeris. The follow-up ephemeris was adjusted to account for the 80 minute offset. The July observation confirmed an on-time arrival of a $\sim$1400 ppm deep full transit relative to the adjusted ephemeris, indicating that the transit timing is consistent with a linear ephemeris. The images have stellar PSF FWHMs of $\sim$2$\farcs0$, and the transit signal is reliably detected on target using a follow-up aperture with radius as small as $1\farcs5$. Therefore, the aperture is negligibly contaminated by the nearest \Gaia\ neighbor $10\farcs5$ south. Systematic effects start to dominate the light curve for smaller apertures. The August B-band observation confirmed an on-time arrival of a $\sim$1400 ppm deep full transit, indicating that the transit-like event does not show a filter dependent depth in B and z-short bands, which photometrically strengthens the case for a transiting exoplanet orbiting around TOI-132.

\section{HARPS Spectroscopic Follow-up}
TOI-132 was observed using HARPS \citep{Pepe2002} spectrograph mounted at the 3.6-m ESO telescope at La Silla observatory, during seven consecutive nights between April 2 and 9 2019, as part of the observing program 0103.C-0442. The exposure time was set to 1200-1800 sec, which allowed us to achieve a mean signal-to-noise (S/N) ratio of $\sim$35 per pixel at 5500\,\AA\ in the extracted spectra. Upon examination of the radial velocities and after performing a 1-planet model fit to the \tess\, period, we found it necessary to acquire more observations to improve the phase coverage. Therefore, 13 additional radial velocities were taken in two runs between May and July 2019, as part of the observing program 1102.C-0923, covering the initial gaps in the orbital phase from the observations in April. We set the exposure time to 1800-2100 sec, leading to a mean S/N ratio of $\sim$40. 

We reduced the spectra using the HARPS online data reduction software (DRS) \citep{Bouchy2001}. The data products include the extracted spectra, both in {\it echelle} and order-merged spectra, the cross-correlation functions\footnote{Obtained using a G2 numerical mask.} (\citealp[CCF,][]{Baranne1996,Pepe2002}) and a measurement of the full-width at half maximum (FWHM) of the CCF profile, and the bisector inverse slope (\citealp[BIS,][]{Queloz2001}). 

\begin{figure*}
    \centering
    \includegraphics[scale=0.485]{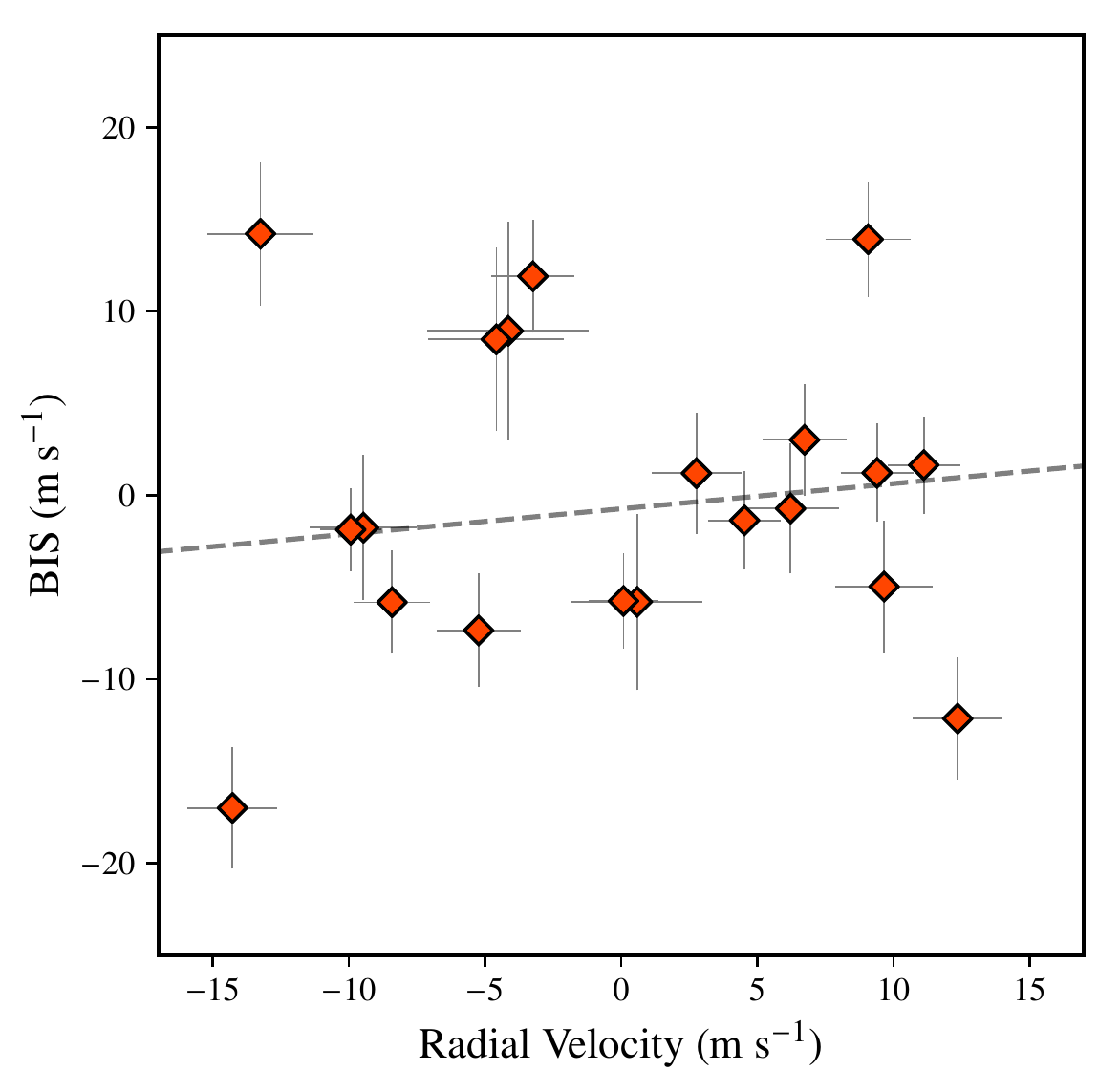}
    \includegraphics[scale=0.485]{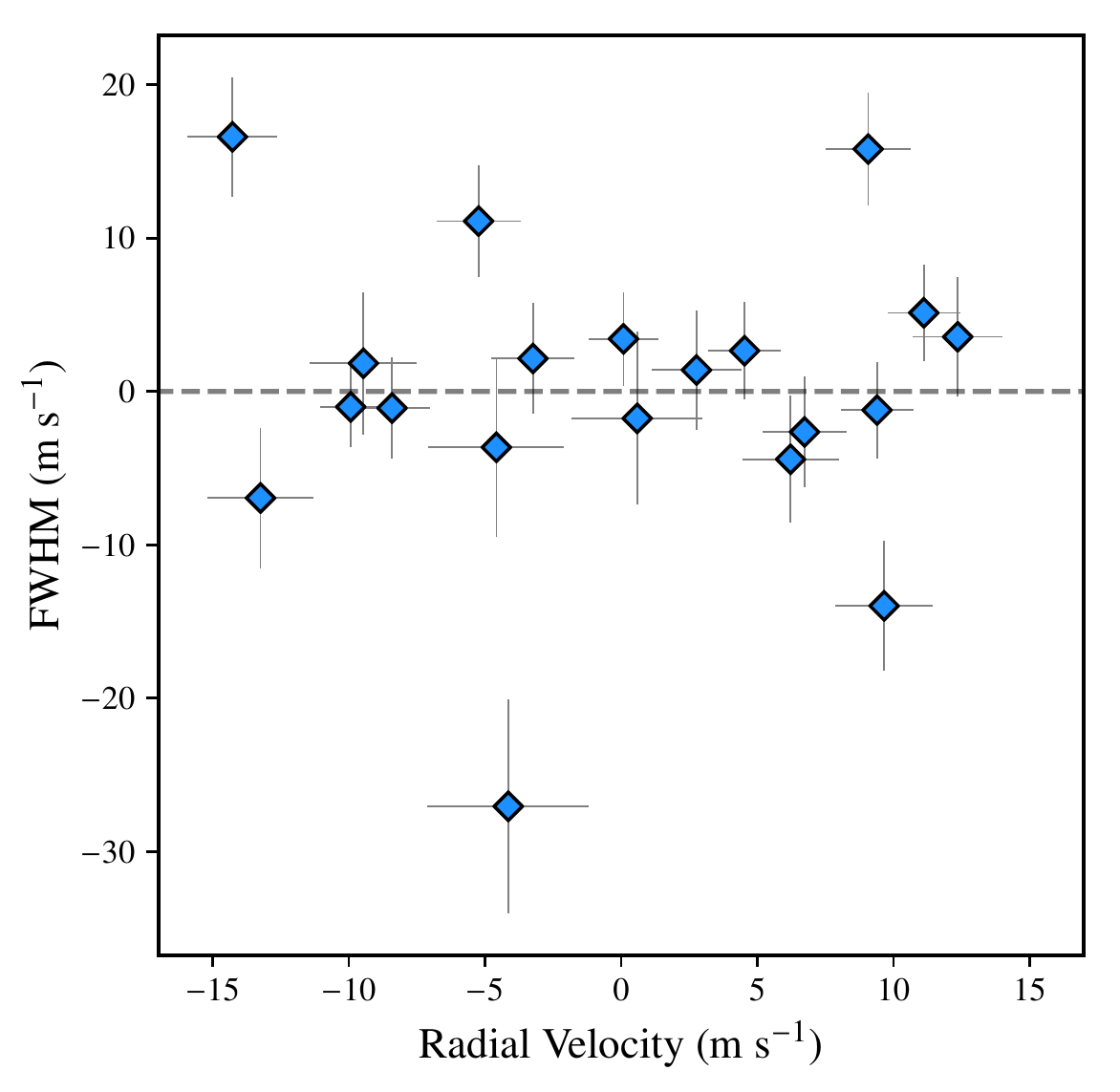}
    \includegraphics[scale=0.485]{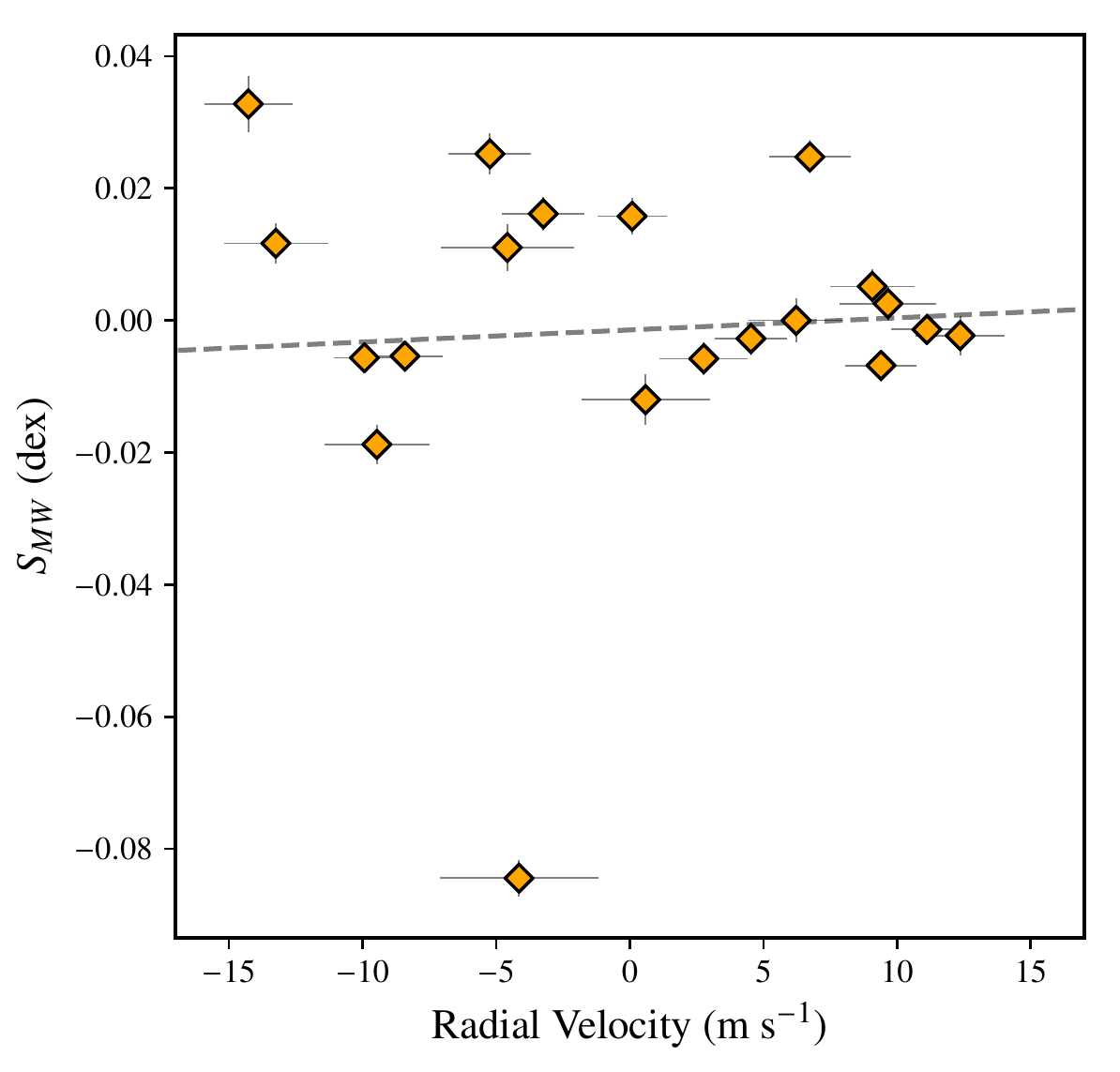}
    \caption{Left to right: correlations between BIS, cross-correlation function FWHM, S-index and radial velocities after subtraction of their mean, respectively. The first two are obtained from DRS and the latter is derived from the HARPS spectra using the HARPS-TERRA algorithm. On each plot, the dashed line represents a linear fit between the activity index and radial velocity. All three plots shows no strong evidence for correlation, although outliers are seen in the FWHM and S$_{MW}$.} 
    \label{fig:correlations}
\end{figure*}

\begin{figure}
\includegraphics[width=\columnwidth]{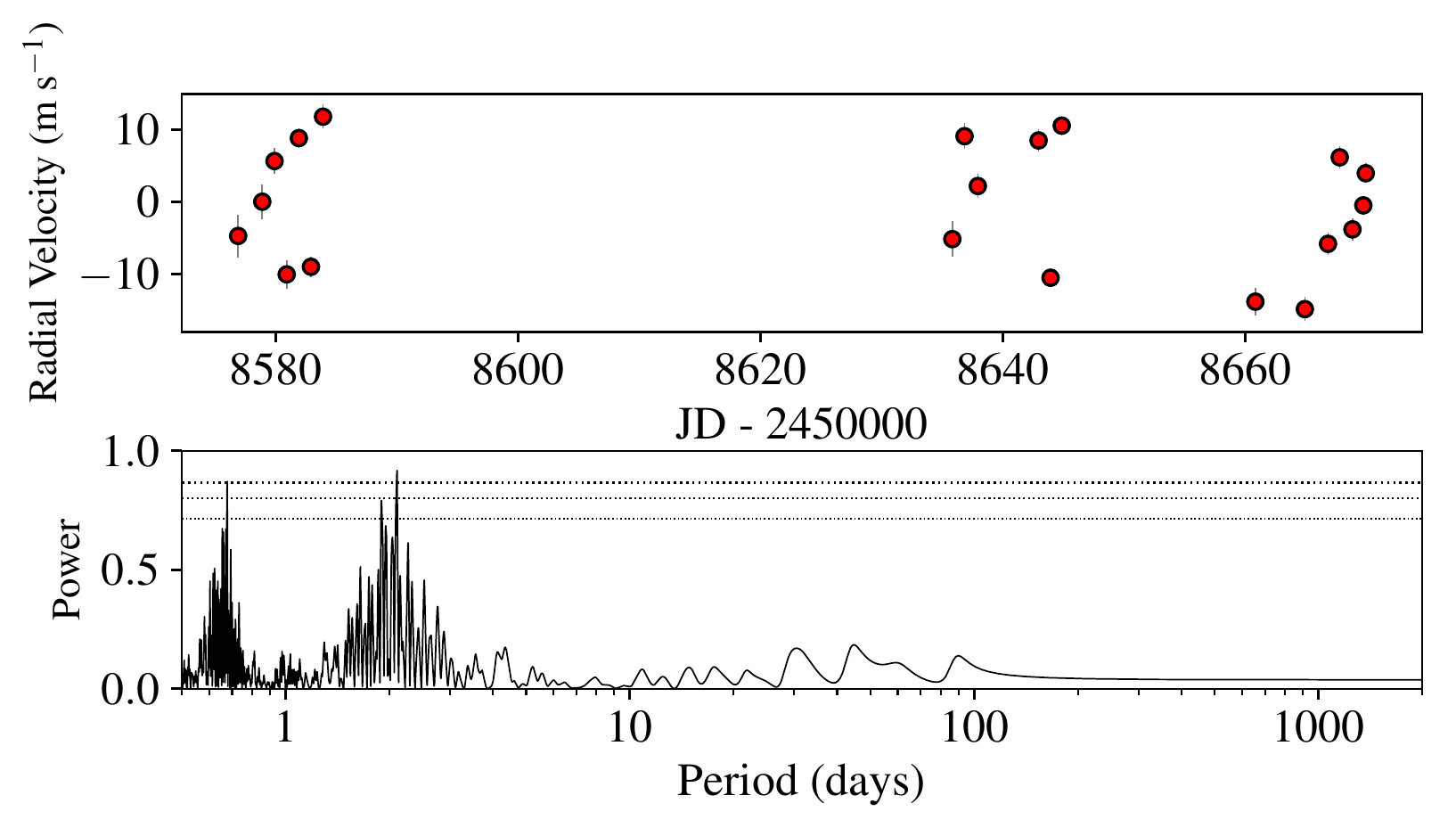}
\caption{Top: Time series showing the radial velocities from the HARPS follow-up observations. Bottom: Generalized Lomb-Scargle periodogram of the radial velocities. Horizontal lines, from bottom to top, represent the 10, 1 and 0.1\% significance levels estimated via 5000 bootstrap samples.}\label{fig:rv_per}
\end{figure}
\begin{figure}
\includegraphics[width=\columnwidth]{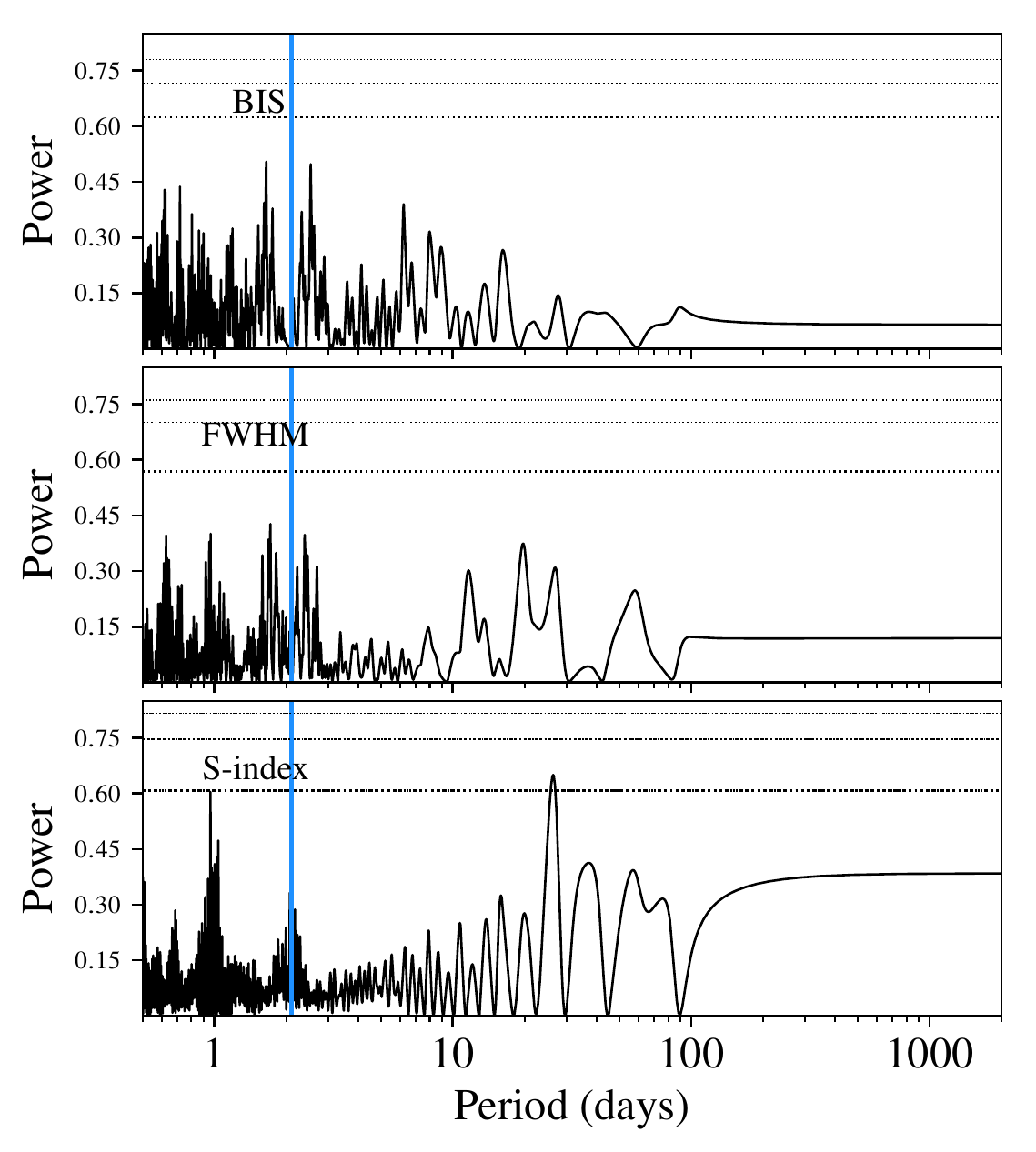}
\caption{Top to bottom: Generalized Lomb-Scargle periodogram for the activity indices obtained with HARPS: BIS, FWHM and S-index, respectively. Horizontal lines, from bottom to top, represent the 10, 1 and 0.1\% significance levels estimated via 5000 bootstrap samples. Vertical line on each plot marks the position of the 2.11-day planet candidate signal present in the radial velocity.}\label{fig:act_per}
\end{figure}

We extracted the radial velocity measurements using the HARPS-TERRA package \citep{AngladaEscude2012}. The algorithm creates a high signal-to-noise template by combining all the observed spectra, based on their signal-to-noise ratio, and then it recomputes the radial velocity of a given observation by matching each individual spectrum with the template. One advantage for choosing HARPS-TERRA is that RVs are computed for every echelle order so it is relatively easy to find the orders with most of the RV information, discarding contaminated or low S/N orders. In this case, we chose the first 22 orders as they produced lowest errors and smallest RMS in the RVs. The software does not compute the BIS nor FWHM of the CCF, which are taken directly from the DRS using a G2 mask. TERRA does compute activity indicators in the form of  S-indices directly from each observed spectrum. The S-index is measured from the cores of the Calcium {\sc ii} H \& K lines ($\lambda_{H}=3933.664\,$\AA, $\lambda_{K}=3968.470\,$\AA) and compared with the flux on adjacent chunks in the continuum, following the prescription from \citet{Lovis2011} and it is calibrated to the Mt. Wilson system ($S_{MW}$), serving as a direct proxy to monitor the chromospheric activity of the star. Uncertainties in BIS are taken as twice the internal RV errors and the FWHM error are 2.35 times the RV uncertainties (see \citealt{Zechmeister2013,Santerne2015}). The results are shown in Table~\ref{tab:rvs}.

\begin{figure} 
\centering
\includegraphics[width=\columnwidth]{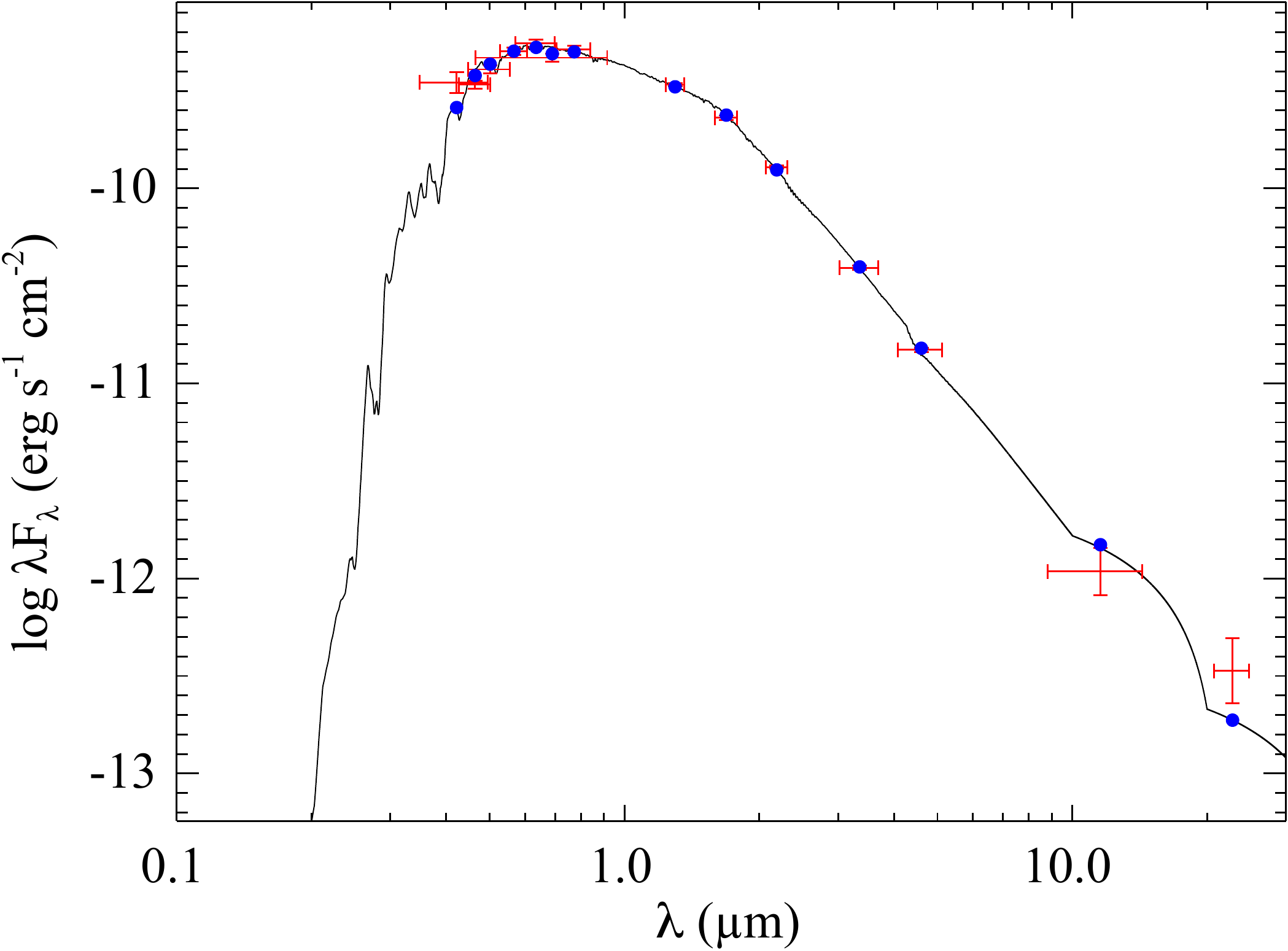}
\caption{Spectral energy distribution of TOI-132. The blue points are the predicted integrated fluxes and the red symbols are the observed values at the corresponding passbands, where the horizontal bars represent the effective width of the passband and the vertical errors represent the $1\sigma$ uncertainties. The best-fit Kurucz atmospheric model for TOI-132 is shown by the black solid line.}
\label{fig:sed}
\end{figure}

Figure \ref{fig:correlations} shows the correlations between radial velocities and activity indicators, BIS, FWHM CCF and $S_{MW}$, from left to right, respectively. No correlations are seen between the radial velocities and the activity indicators. However, we note one outlier point in the FWHM and S-index, which was related with an observation acquired under poor weather conditions at the beginning of the observing run in April 2019.

We computed the Generalized Lomb-Scargle periodogram\footnote{\texttt{ astropy.timeseries.LombScargle()},\url{https://docs.astropy.org/en/stable/timeseries/lombscargle.html}.} (\citealp[GLS;][]{Zechmeister2009}) of the HARPS Doppler measurements and activity indicators. As shown in Figure~\ref{fig:rv_per}, the GLS periodogram of the HARPS RVs shows a significant peak at the orbital period of the transiting planet (2.11 d) with a false-alarm probability  FAP\,$<$\,0.1\%. We note that the secondary peak with FAP\,$<$\,1\% is the 1-day alias of the orbital period. The periodograms of the HARPS activity indicators show neither a significant peak matching the one found in the RVs, nor any other significant peaks  (Figure \ref{fig:act_per}).

\begin{table*}
	\centering
	\caption{HARPS Radial Velocities and spectral activity indices for TOI-132}
	\begin{tabular}{lcccccccc} 
\hline\hline
	BJD 	&	RV & $\sigma$RV & $S_{MW}$ &$\sigma S_{MW}$ &FWHM &$\sigma$FWHM& BIS &$\sigma$BIS\\
	\hline\hline
	(- 2450000)& (\mps) & (\mps)& (dex) & (dex) &(\kps)& (\mps)&(\mps) &(\mps)\\
	\hline
8576.90725 & -4.737 & 2.967 & 0.056 & 0.003 & 6.885 & 16.180 & 2.967 & 5.933 \\
8578.89655 & 0.000 & 2.398 & 0.128 & 0.004 & 6.911 & 16.240 & 2.398 & 4.797 \\
8579.90764 & 5.631 & 1.765 & 0.140 & 0.003 & 6.908 & 16.234 & 1.765 & 3.531 \\
8580.90988 & -10.056 & 1.972 & 0.121 & 0.003 & 6.914 & 16.248 & 1.972 & 3.943 \\
8581.91433 & 8.808 & 1.338 & 0.133 & 0.002 & 6.911 & 16.241 & 1.338 & 2.675 \\
8582.91045 & -9.005 & 1.402 & 0.135 & 0.002 & 6.911 & 16.241 & 1.402 & 2.803 \\
8583.90870 & 11.771 & 1.656 & 0.138 & 0.003 & 6.916 & 16.252 & 1.656 & 3.312 \\
8635.81477 & -5.174 & 2.488 & 0.151 & 0.004 & 6.909 & 16.235 & 2.488 & 4.977 \\
8636.82174 & 9.069 & 1.800 & 0.143 & 0.003 & 6.898 & 16.211 & 1.800 & 3.599 \\
8637.91868 & 2.175 & 1.649 & 0.134 & 0.003 & 6.914 & 16.247 & 1.649 & 3.297 \\
8642.93057 & 8.481 & 1.571 & 0.145 & 0.003 & 6.928 & 16.281 & 1.571 & 3.142 \\
8643.91730 & -10.522 & 1.129 & 0.134 & 0.002 & 6.911 & 16.242 & 1.129 & 2.257 \\
8644.84072 & 10.526 & 1.331 & 0.139 & 0.002 & 6.917 & 16.256 & 1.331 & 2.662 \\
8660.81222 & -13.834 & 1.945 & 0.152 & 0.003 & 6.905 & 16.228 & 1.945 & 3.891 \\
8664.89377 & -14.864 & 1.652 & 0.173 & 0.004 & 6.929 & 16.283 & 1.652 & 3.305 \\
8666.80357 & -5.826 & 1.542 & 0.165 & 0.003 & 6.923 & 16.270 & 1.542 & 3.084 \\
8667.76863 & 6.145 & 1.530 & 0.165 & 0.003 & 6.910 & 16.238 & 1.530 & 3.061 \\
8668.82036 & -3.829 & 1.534 & 0.156 & 0.003 & 6.914 & 16.249 & 1.534 & 3.067 \\
8669.71698 & -0.505 & 1.294 & 0.156 & 0.003 & 6.916 & 16.252 & 1.294 & 2.588 \\
8669.91776 & 3.943 & 1.344 & 0.137 & 0.003 & 6.915 & 16.250 & 1.344 & 2.687 \\
	\hline
\end{tabular}
    \label{tab:rvs}
\end{table*}

\begin{table*}
    \centering
    \caption{Below are the priors used for TOI-132 for the \textit{final} joint analysis fit using \juliet.
    As a reminder, $p=R_p/R_*$ and $b=(a/R_*)\cos(i_p)$, where $R_p$ is the planetary radius, $R_*$ the stellar radius, $a$ the semi-major axis of the orbit and $i_p$ the inclination of the planetary orbit with respect to the plane of the sky. $e$ and $\omega$ are the eccentricity and argument of periastron of the orbits. The prior labels of $\mathcal{N}$, $\mathcal{\tilde{N}}$, $\mathcal{U}$, and $\mathcal{J}$ represent normal, truncated normal, uniform, and Jeffreys distributions. See text for explanations about other parameters.}
    \label{tab:priors}
    \begin{tabular}{lccl} 
        \hline
        \hline
        Parameter name & Prior & Units & Description \\
                \hline
                \hline        
        Parameters for planet b\\
        ~~~$P_b$ & $\mathcal{N}(2.10937,0.001)$ & days & Period. \\
        ~~~$T_{0,b} - 2458000$ & $\mathcal{N}(337.451,10)$  & days & Time of transit-center. \\
        ~~~$r_{1,b}$ & $\mathcal{U}(0,1)$ & --- & Parametrization for $p$ and $b$${}^{1}$. \\
        ~~~$r_{2,b}$ & $\mathcal{U}(0,1)$ & --- & Parametrization for $p$ and $b$${}^{1}$. \\
        ~~~$a_{b}$ & $\mathcal{U}(4.5,7.0)$ & ---& scaled semi-major axis.\\
        ~~~$K_{b}$ & $\mathcal{N}(12.1,2.0)$ & m s$^{-1}$ & Radial-velocity semi-amplitude. \\
       ~~~${e}_{b}$ & $\mathcal{\tilde{N}}(0,0.1,0,0.25) $ & ---& eccentricity.\\
       ~~~$\omega_{b}$& $\mathcal{U}(0,359.)$ & degrees & argument of periastron.\\
        
        \hline
        Parameters for TESS\\
        ~~~$D_{\textnormal{TESS}}$ &  1.0 (Fixed)  & --- & Dilution factor for \textit{TESS}. \\
        ~~~$M_{\textnormal{TESS}}$ & $\mathcal{N}(0,1)$ & ppm & Relative flux offset for \textit{TESS}. \\
        ~~~$\sigma_{w,\textnormal{TESS}}$ & $\mathcal{J}(0.1,100)$ & ppm & Extra jitter term for \textit{TESS} lightcurve. \\
        ~~~$q_{1,\textnormal{TESS}}$ & $\mathcal{U}(0,1)$ & --- & Quadratic limb-darkening parametrization. \\
        ~~~$q_{2,\textnormal{TESS}}$ & $\mathcal{U}(0,1)$ & --- & Quadratic limb-darkening parametrization. \\
        \hline
        Parameters for LCO\\
        ~~~$D_{\textnormal{LCO}}$ & 1.0 (Fixed) & --- & Dilution factor for LCO. \\
        ~~~$M_{\textnormal{LCO}}$ & $\mathcal{N}(0,1)$ & ppm & Relative flux offset for LCO.\\
        ~~~$\sigma_{w,\textnormal{LCO}}$ & $\mathcal{J}(0.1,100)$ & ppm & Extra jitter term for LCO lightcurve. \\
        ~~~$q_{1,\textnormal{LCO}}$ & $\mathcal{U}(0,1)$ & --- & Quadratic limb-darkening parametrization. \\
        ~~~$q_{2,\textnormal{LCO}}$ & $\mathcal{U}(0,1)$ & --- & Quadratic limb-darkening parametrization. \\
        \hline
        Parameters for HARPS\\
        ~~~$\mu_{\textnormal{HARPS}}$ & $\mathcal{N}(-0.6,1.)$ &m s$^{-1}$ & Radial velocity zero-point (offset).\\
        ~~~$\sigma_{w,\textnormal{HARPS}}$ & $\mathcal{J}(0.1,10)$ & m s$^{-1}$ & Extra jitter term for HARPS radial velocities. \\
        \hline
       \multicolumn{3}{l}{$^{1}$We used the transformations outlined in \citet{Espinoza2018efficient} and also set $p_{l}$=0.03 and $p_{u}$=0.05 in the \texttt{juliet} call.}\\
    \end{tabular}
\end{table*}

\section{Stellar Parameters}\label{sec:star}
We first estimated the stellar parameters by combining the HARPS spectra into a high-S/N ratio spectrum and fed that into the \texttt{SPECIES} code \citep{Soto2018}. For a more detailed explanation and outputs from this code, the reader is referred to \cite{Diaz2018} and \cite{Soto2018}.

We also analyzed the combined HARPS spectrum using both Spectroscopy Made Easy \citep[\texttt{SME}, version 5.22;][]{Valenti1996, Valenti2005, Piskunov2017}, and the empirical package \texttt{SpecMatch-Emp} \citep{Yee2017}. We followed the same procedures outlined in, e.g., \citet{Fridlund2017}, \citet{Persson2018,Persson2019}, \citet{Gandolfi2019}. The two methods provide consistent results within 1--2\,$\sigma$, which are also in agreement with those obtained with \texttt{SPECIES}. In particular, the age of the star was determined by isochrone fitting according to the method described in \texttt{SPECIES}. We note that, while there is no reason to prefer one set of spectroscopic parameter estimates over the others, we adopted the results derived with \texttt{SPECIES} for the subsequent analyses presented in this work. 

We performed an analysis of the broadband spectral energy distribution (SED) of the star together with the {\it Gaia\/} DR2 parallaxes \citep[adjusted by $+0.08$~$mas$ to account for the systematic offset reported by][]{StassunTorres2018}, in order to determine an empirical measurement of the stellar radius, following the procedures described in \citet{StassunTorres2016} and \citet{Stassun2017,Stassun2018a}. We retrieved the NUV flux from {\it GALEX}, the $B_T V_T$ magnitudes from {\it Tycho-2}, 
the $BVgri$ magnitudes from APASS, the $JHK_S$ magnitudes from {\it 2MASS}, the W1--W4 magnitudes from {\it WISE}, and the $G$ magnitude from {\it Gaia}. Together, the available photometry spans the full stellar SED over the wavelength range 0.2--22~$\mu$m (see Figure~\ref{fig:sed}).  \newline
We performed a fit using Kurucz stellar atmosphere models, with the effective temperature ($T_{\rm eff}$) and metallicity ([Fe/H]) and surface gravity ($\log g$) adopted from the spectroscopic analysis of \texttt{SPECIES}. The only free parameter is the extinction ($A_V$), which we restricted to the maximum line-of-sight value from the dust maps of \citet{Schlegel1998}. The resulting fit shown in Figure~\ref{fig:sed}, gives a reduced $\chi^2$ of 2.4 and best-fit $A_V = 0.03 \pm 0.01$. Integrating the (unreddened) model SED, it gives the bolometric flux at Earth, $F_{\rm bol} = 7.492\pm0.087 \times 10^{-10}$ erg\,s$^{-1}$\,cm$^{-2}$. Taking the $F_{\rm bol}$ and $T_{\rm eff}$ together with the {\it Gaia\/} DR2 parallax, gives the stellar radius, $R_\star=0.90\pm0.02\,R_\odot$. Finally, we can use the empirical relations of \citet{Torres2010} and a 6\% error from the empirical relation itself to estimate the stellar mass, $M_\star=0.97\pm0.06\,M_\odot$; this, in turn, together with the stellar radius provides an empirical estimate of the mean stellar density $\rho_\star = 1.89 \pm 0.15$ g~cm$^{-3}$. We note the small errorbars on both stellar mass and radius come directly from propagation of uncertainties in $T_{\rm eff}$, $F_{\rm bol}$, and parallax. In this case, the fractional errors are of order $\sim$1\%, $\sim$~1\% and $\sim$0.5\%, respectively. Then, the uncertainty in stellar radius is dominated by the $T_{\rm eff}$ error, in this case that implies an error of $\sim$2\% (see Table \ref{tab:star}). 

\begin{figure} 
    \centering
    \includegraphics[width=\columnwidth]{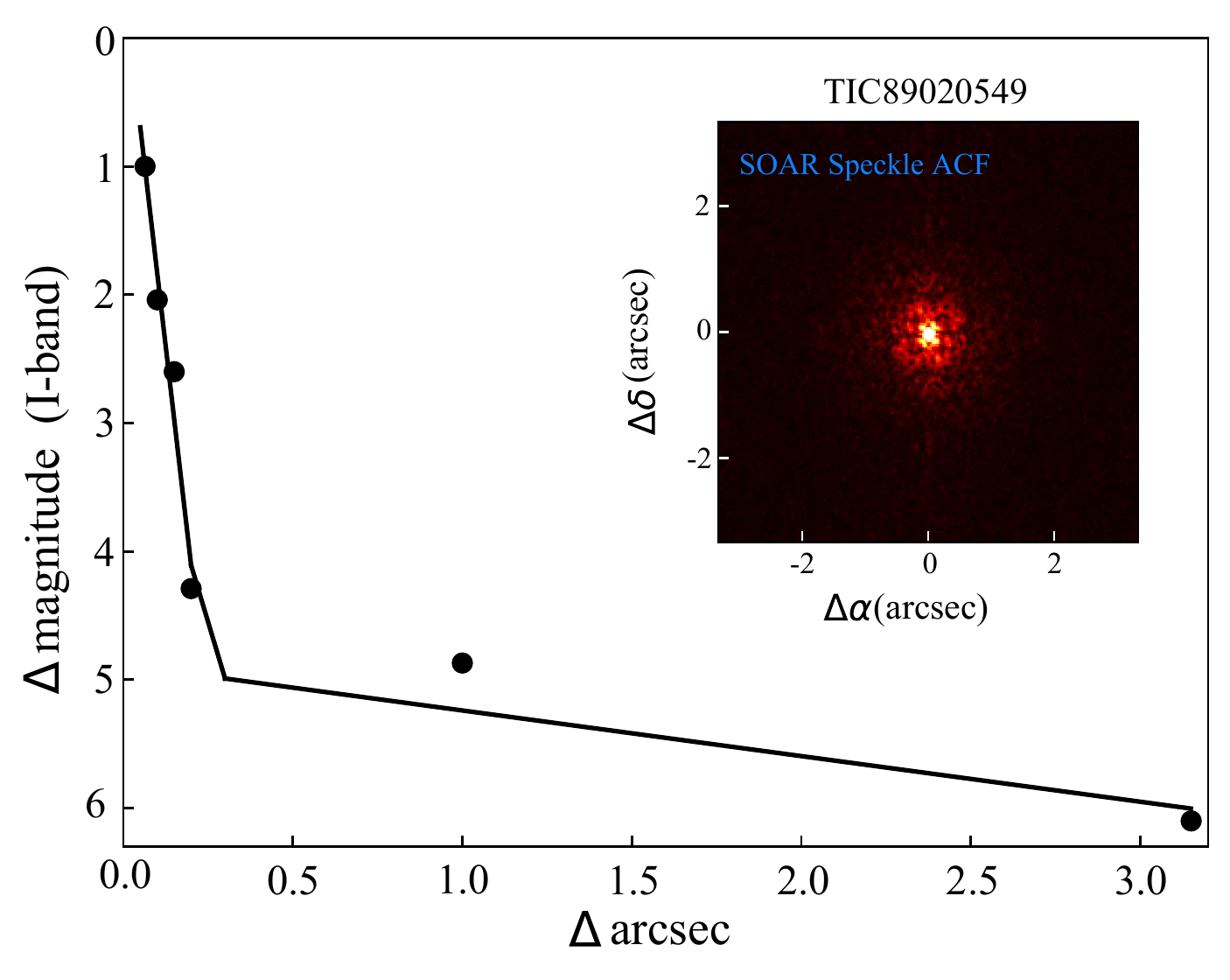}
    \caption{Speckle imaging for TOI-132 obtained with SOAR. Inset on the top right corner shows a preview of the ACF. }
    \label{fig:speckle}
\end{figure}

\section{Speckle Imaging}
The relatively large 21-arcsec pixels of \tess\, can result in contamination from companion stars or nearby sources. The additional light from these can dilute the planetary transit, resulting in an underestimated planet radius. We searched for nearby sources with speckle imaging with HRCam on the 4.1-m Southern Astrophysical Research (SOAR) telescope \citep{Tokovinin2018} on 2018 September 25 UT. From these observations, a potential companion star was detected at low-significance. The purported star was located near the first diffraction ring of the primary star, at 0.079 arcsec (and a projected distance of $\sim$12 AU), a similar position as optical ghosts which can occasionally appear in the speckle imaging during periods of low wind. This triggered a warning as the flux contamination due to the companion ($\Delta$m$\sim$2.6 mag) would have not been negligible for the spectroscopic observations given that the diameter of the fibers on HARPS is $\sim$1 arcsec, meaning that the suspected companion was inside the aperture of the fiber. Upon visual inspection of the CCF and the individual spectra, we could not see evidence for such a contamination. The system was observed again on 2019 May 18 UT in excellent conditions, and the possible companion star was not detected. The 5-$\sigma$ detection sensitivity and auto-correlation function of the later observation are shown in Figure \ref{fig:speckle}. 

\section{ASAS Photometry}
\begin{figure}
    \centering
    \includegraphics[width=\columnwidth]{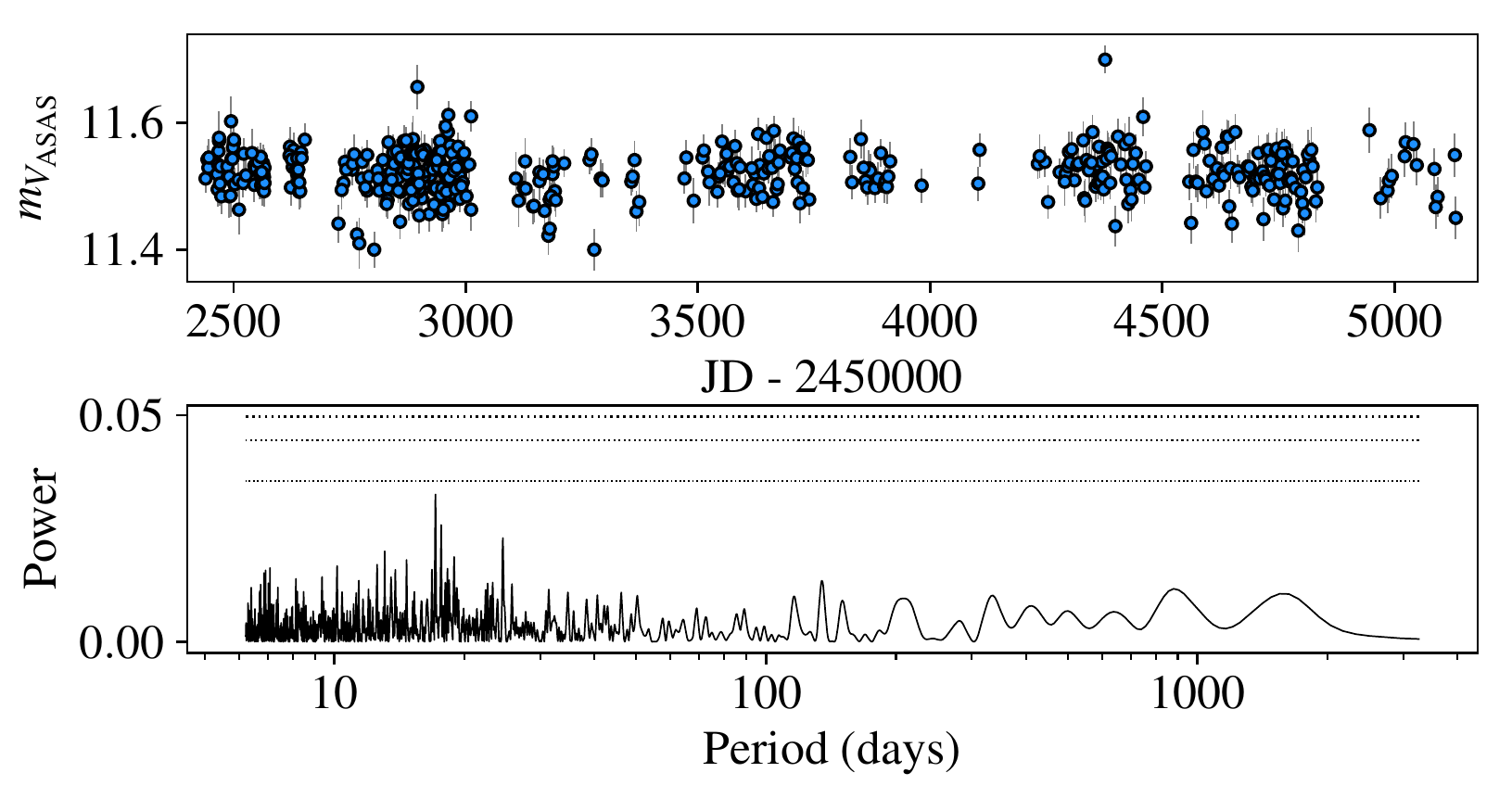}  
    \caption{ASAS V-band photometry for TOI-132 to search for additional sources of periodicity in the star. The bottom plot shows the Generalized Lomb-Scargle periodogram of the time series. Horizontal lines, from bottom to top, represent the 10, 1 and 0.1\% significance levels estimated via 5000 bootstrap samples.}
    \label{fig:asas}
\end{figure}

We analyzed photometry from the All-Sky Automated Survey (\citealp[ASAS,][]{Pojmanski1997}) to search for stellar rotational periods. There are 694 available photometric measurements spanning 8.9 years, from November 2000 to December 2009. The selection of the best aperture was made choosing the time series with the lowest Median Absolute Deviation (MAD)\footnotetext{MAD = median$(|X_{i}  -\bar{X}|)$/0.6745 }. We discarded 129 points that were flagged as bad datapoints, including only 565 measurements with either ``A'' or ``B'' quality. Figure \ref{fig:asas} shows the photometric time series after removing outliers and bad data and the GLS. 

From the power spectrum in the periodogram the highest power is found to be at 17.138 days. We estimated the 10, 1 and 0.1\% significance level by running 5000 bootstrap samplings using the implementation available in the \texttt{Python} module \texttt{astropy.stats.false\_alarm\_probability()}\footnote{\url{https://docs.astropy.org/en/stable/timeseries/lombscargle.html}}. Although the highest peak in the GLS periodogram is noticeable and unique, its significance is below the 10\% level, as seen from the bottom panel in the figure.

\begin{figure*}
	\centering
	    \includegraphics[scale=0.65]{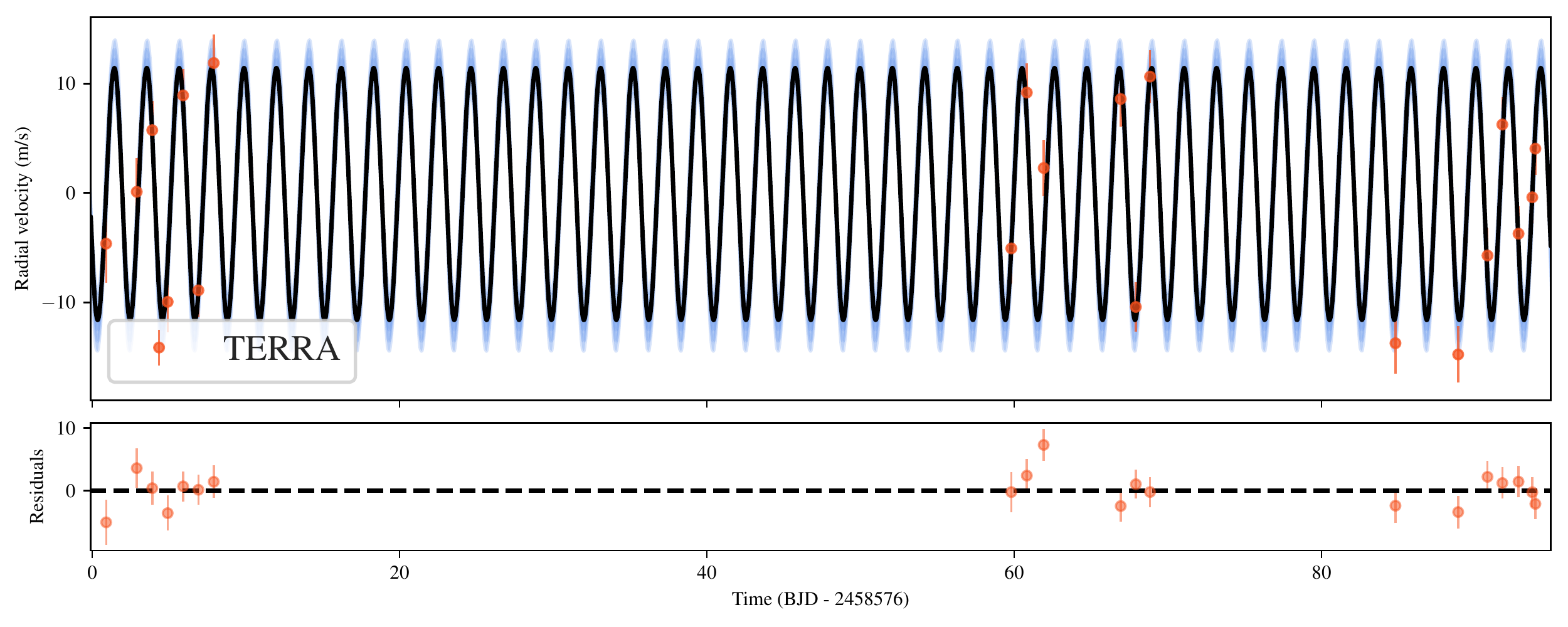}
    \includegraphics[scale=0.56]{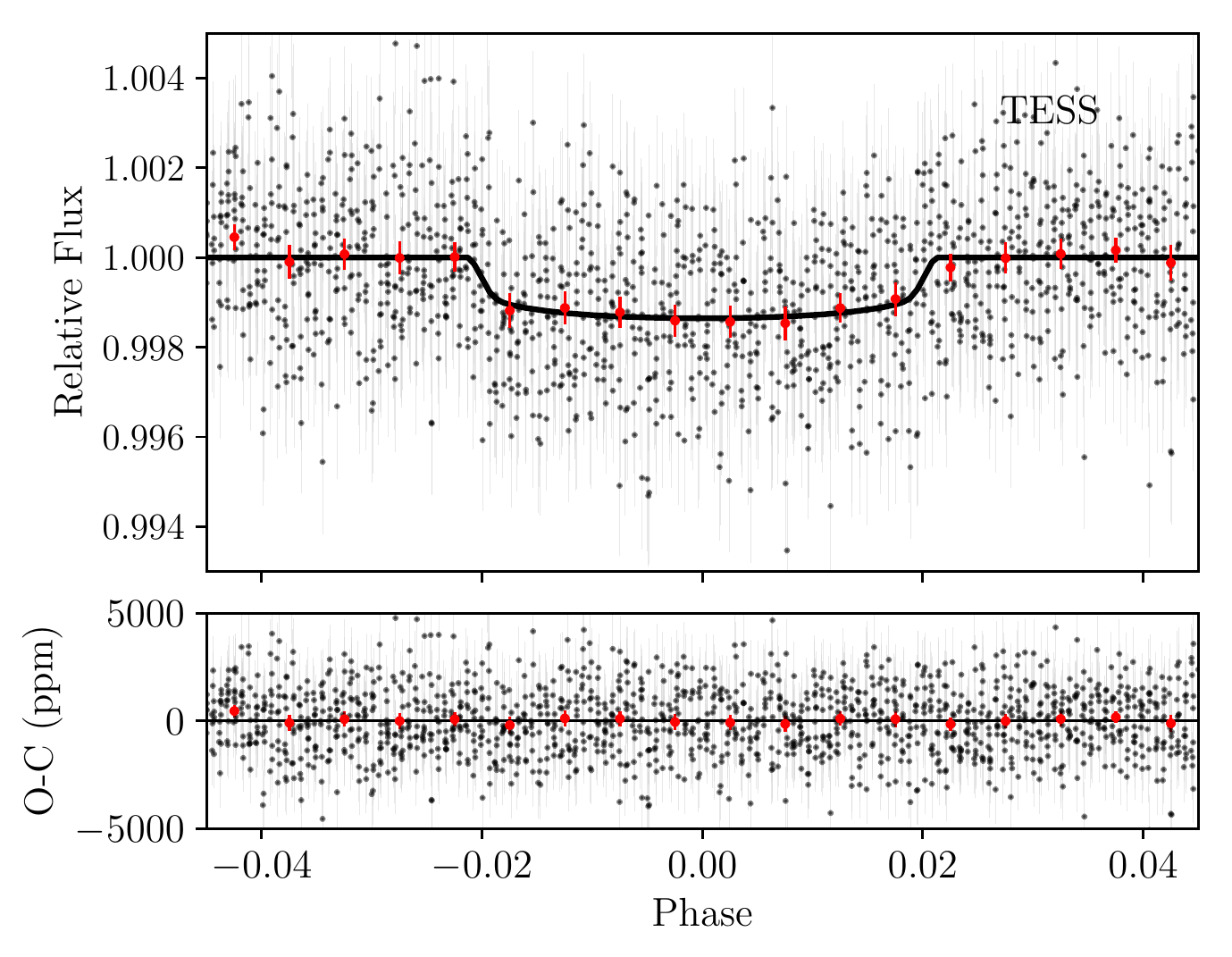}
    \includegraphics[scale=0.56]{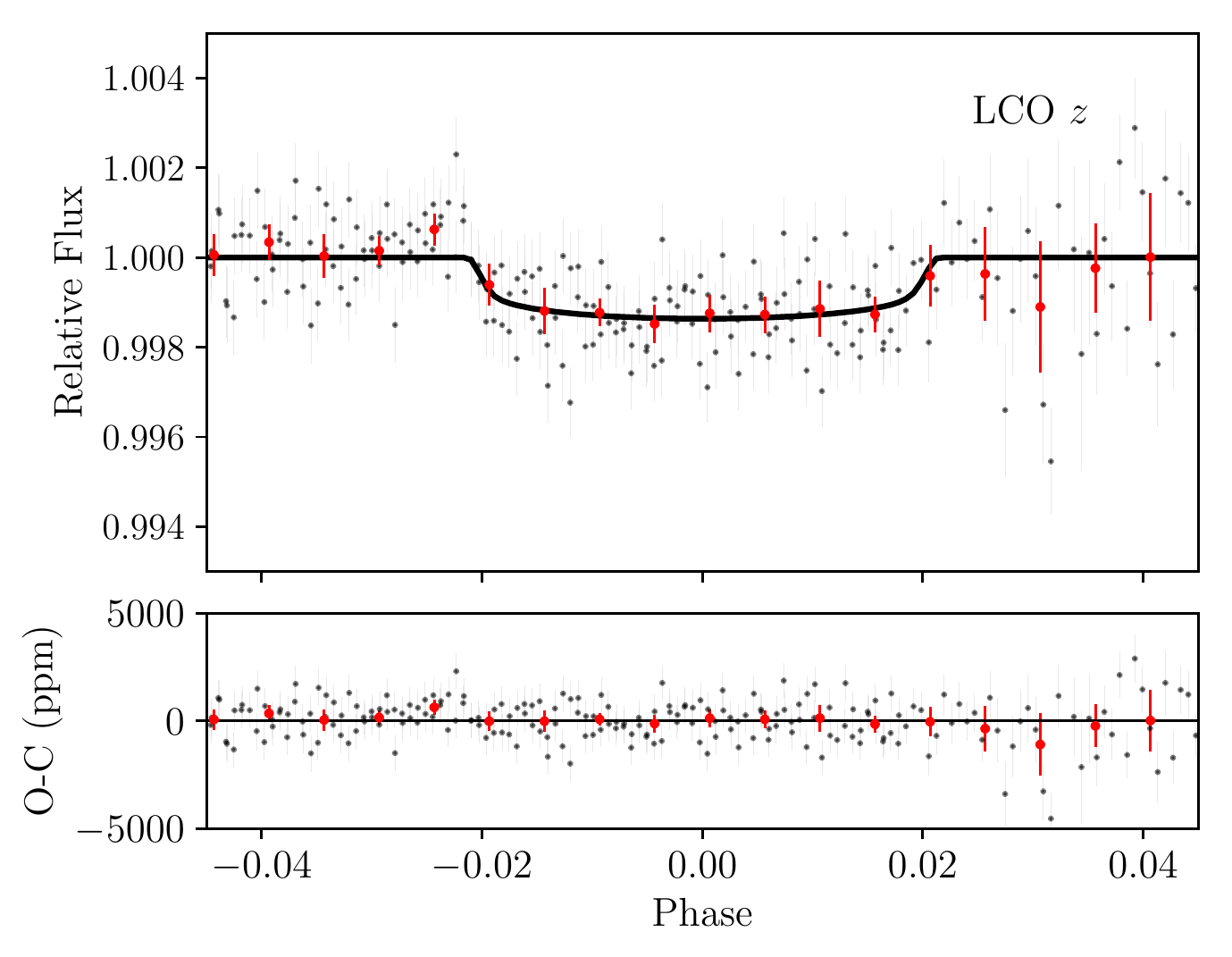}
  	\includegraphics[width=\columnwidth]{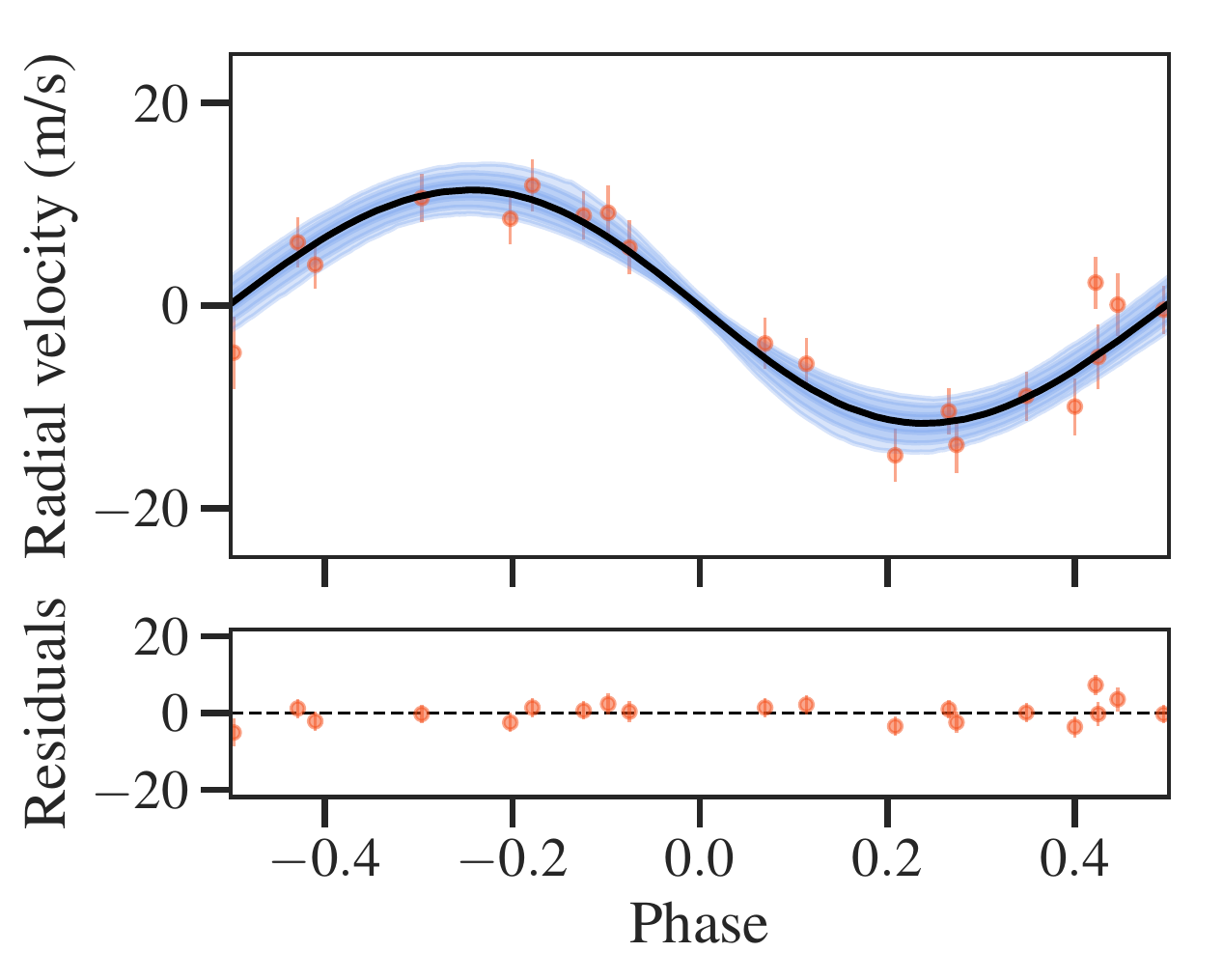}
    \caption{Results from the joint fit for the 1-planet model. {\it Top panel:} HARPS-TERRA radial velocities and best-fit Keplerian model (solid curve) the bands around it show 68\%, 95\% and 99\% posterior credibility bands. {\it Mid panels:} {\it TESS} photometry (left) and LCO $z$-short photometry (right) phase-folded to the 2.109 d period of \planet along with best-fit transit model from the joint fit. Red points show the binned photometry in phase bins of 0.005. {\it Bottom panel:} phase-folded RVs from HARPS. The black line shows the model. Credibility bands are shown in the same way as in top panel. The error bars of both photometry and RV data include their corresponding jitter. }
    \label{fig:joint_fit_phased}
\end{figure*}

\section{Joint Analysis}
We modeled the radial velocities and the photometry using the \juliet\footnote{\url{https://github.com/nespinoza/juliet}} package \citep{Espinoza2019b}. Table \ref{tab:priors} shows the priors used in the analysis. 
We set up the initial priors for the period of the candidate and the time of transit ($T_{0}$) using the reported values in the \tess\, DV report document for TOI-132. 

Preliminary analysis was done by making use of \texttt{Systemic Console v2} \citep{Meschiari2009}. We analyzed the radial velocities only to get an initial rough estimate of both instrumental and orbital parameters of the system such as the velocity semi-amplitude, eccentricity and minimum mass of the planet. The period and transit time were constrained using the updated values provided by \tess. Initial results for a 1-planet model with eccentricity fixed at zero, yields an RMS$\sim$2.7 \mps. Letting eccentricity and argument of periastron as free parameters the best-fit model RMS goes down to $\sim$2.5 \mps\, and e$\sim$0.17. With this information from the \texttt{Systemic} analysis we set up the priors on two runs with \texttt{juliet}. This package has been proven to be an excellent tool for analyzing both photometry and radial velocities using a joint model (see e.g., \citealp{Brahm2019, Espinoza2019a, Kossakowski2019}). In short, the code uses \texttt{batman} \citep{Kreidberg2015} to model the transit data and \texttt{radvel} \citep{Fulton2018} to model the radial velocities, and in order to estimate the Bayesian log-evidence, ln$Z$, for model comparison we used the option of the Dynamic Nested Sampling algorithm that the \texttt{dynesty} \citep{Speagle2018,Speagle2019} package provides. We note that, while \texttt{juliet} has the option to include Gaussian Processes to model the lightcurve, radial velocities or both, we did not set this option as there was no evidence of additional variability in the \texttt{PDCSAP\_FLUX}-corrected lightcurve (see Figure \ref{fig:tesslc}). 
We also used the parametrization described in \citet{Espinoza2018efficient} that allows an efficient way to sample the impact parameter, $b$, and the planet-to-star radius ratio, $p$, where only values that are physically plausible in the ($p,b)$ plane are sampled via the $r_{2}$ and $r_{2}$ coefficients \citep{Espinoza2018efficient}. For the limb-darkening coefficients, we use the parametrization of \citet{Kipping2013} for two parameter laws. All the prior information is listed in Table \ref{tab:priors}.

We set up two different runs, first by fixing eccentricity to zero, and another treating it (along with $\omega$) as free parameter. Comparing the evidences from the circular ($lnZ$=89705.63) and eccentric model ($lnZ$=89709.05) we obtain $\Delta$ ln$Z=3.41$ which suggests moderate evidence the latter is preferred over the circular one according to the model selections criteria and thresholds described in \citet{Espinoza2019b}. 
The joint model results are shown in Figure \ref{fig:joint_fit_phased}. 

As a sanity check, we also performed an independent joint analysis using the \texttt{Python/FORTRAN} software suite \texttt{pyaneti} \citep{Barragan2019a}. Results are consistent with those obtained with \texttt{juliet} well within the nominal error bars. 

Using the luminosity of the host star, we could retrieve the incident flux on \planet by using the planet radius and semi-major axis from our joint model. We estimated that the insolation of \planet is $S_{\rm p}= 860\, S_{\oplus}$.

In order to estimate the average equilibrium temperature of the planet, considering the physical properties of  \planet we assumed a Bond albedo of $A_{B}=0.31$, that corresponds to the value accepted for Neptune. Then

\begin{equation}
T_{\rm eq} = T_{*} \sqrt{\frac{R_{*}}{2a}}(1-A_{B})^{\frac{1}{4}}
\end{equation}
yields an equilibrium temperature of $T_{\rm eq}$=1384$^{+53}_{-75} \, K$ for the planet.

\begin{table}
	\centering
	\caption{Planetary Properties for TOI-132\,b}
	\begin{tabular}{lc} 
\hline\hline
	Property	&	Value \\
	\hline\hline
    Fitted Parameters &\\
    $P$ (days)		&	2.1097008$^{+0.000012}_{-0.000011}$  \\\vspace{2pt}
    $T_0$ (BJD - 2450000)&	8343.77954$^{+0.00093}_{-0.00092}$	\\\vspace{1pt}
    $a/R_{*}$		&6.325$^{+0.440}_{-0.657}$ \\\vspace{2pt}
    $b$ & 0.583$^{+0.107}_{-0.149}$    \\\vspace{2pt}
	$K$ (\ms) 	&11.58 $^{+0.78}_{-0.77}$	\\
    $i_{\rm p}$ (deg)  & 84.63$^{+1.58}_{-5.63}$ \\\vspace{2pt} 
    $e$ 			&0.087$^{+0.054}_{-0.057}$ \\
    $\omega$ (deg) & 125.47$^{+109.89}_{-53.77}$\\\hline
    Derived Parameters & \\
    $M_{\rm p}$ (\me)& 22.83$^{+1.81}_{-1.80}$	\\\vspace{2pt}
    $R_{\rm p}$ (\re)& 3.43$^{+0.13}_{-0.14}$ \\\vspace{2pt}
    $a$ (AU) & 0.026$^{+0.002}_{-0.003}$ \\\vspace{2pt}
    $\rho_{\rm p}$ (\gcm) & 3.11$^{+0.44}_{-0.45}$ \\\vspace{2pt}
    $T_{\rm eq}$ (K) &1384$^{+53}_{-75}$ \\
    \hline
    Instrumental Parameters &\\
    $M_{\textnormal{TESS}}$ (ppm) & -0.000068$^{+0.000013}_{-0.000012}$\\\vspace{2pt}
    $\sigma_{w, \rm TESS}$ (ppm) & 8.01$^{+23.29}_{-5.83}$\\\vspace{2pt}
    $q_{1,\rm TESS}$ & 0.296$^{+0.321}_{-0.188}$ \\\vspace{2pt}
    $q_{2,\rm TESS}$ & 0.345$^{+0.342}_{-0.221}$\\\vspace{2pt}
    $M_{\textnormal{LCO}}$ (ppm) & -0.000057$^{+0.000059}_{-0.000058}$ \\\vspace{2pt}
    $\sigma_{w, \rm LCO}$ (ppm)&  461.28$^{+73.22}_{-73.90}$\\\vspace{2pt}
    $q_{1,\rm LCO}$ & 0.374$^{+0.304}_{-0.230}$ \\\vspace{2pt}
    $q_{2,\rm LCO}$ & 0.321$^{0.323}_{-0.208}$\\\vspace{2pt}
    $\mu_{\rm HARPS}$ (m s$^{-1}$)& -0.11$^{+0.51}_{-0.52}$ \\\vspace{2pt}
    $\sigma_{w,\rm HARPS}$ (m s$^{-1}$)&1.99$^{+0.71}_{-0.57}$ \\
    \hline\hline
	\end{tabular}
    \label{tab:planet}
\end{table}

\section{TTV Analysis}
In order to search for possible Transit Timing Variations in TOI-132\,b, we computed the individual transit time of each light curve using the \texttt{EXOFASTv2} code \citep{Eastman2013,Eastman2017}. \texttt{EXOFASTv2} uses the Differential Evolution Markov chain Monte Carlo method (DE-MCMC) to derive the values and their uncertainties for the stellar, orbital and physical parameters of the system.

So as to obtain the transit time of each light curve, we fixed the stellar and orbital parameters to the values obtained from the global fit performed by \texttt{juliet}, except for the transit time and their baseline flux. If a planet follows strictly a Keplerian orbit, the transit time of a given epoch $T_{c}(E)$ is a linear function of the orbital period $P$:

\begin{equation}
   T_{c}(E) = T_{c}(0) + P\times E
\label{eq:eq_transit}
\end{equation}

Where $T_{c}(0)$ is a reference transit time and $E$ is the number of epochs since $T_{c}(0)$. The best-fit values for equation \ref{eq:eq_transit} from \texttt{juliet} are shown in Table \ref{tab:planet} along with the planetary parameters fixed to compute the individual transit time.

\begin{figure}
\centering
\includegraphics[width=\columnwidth]{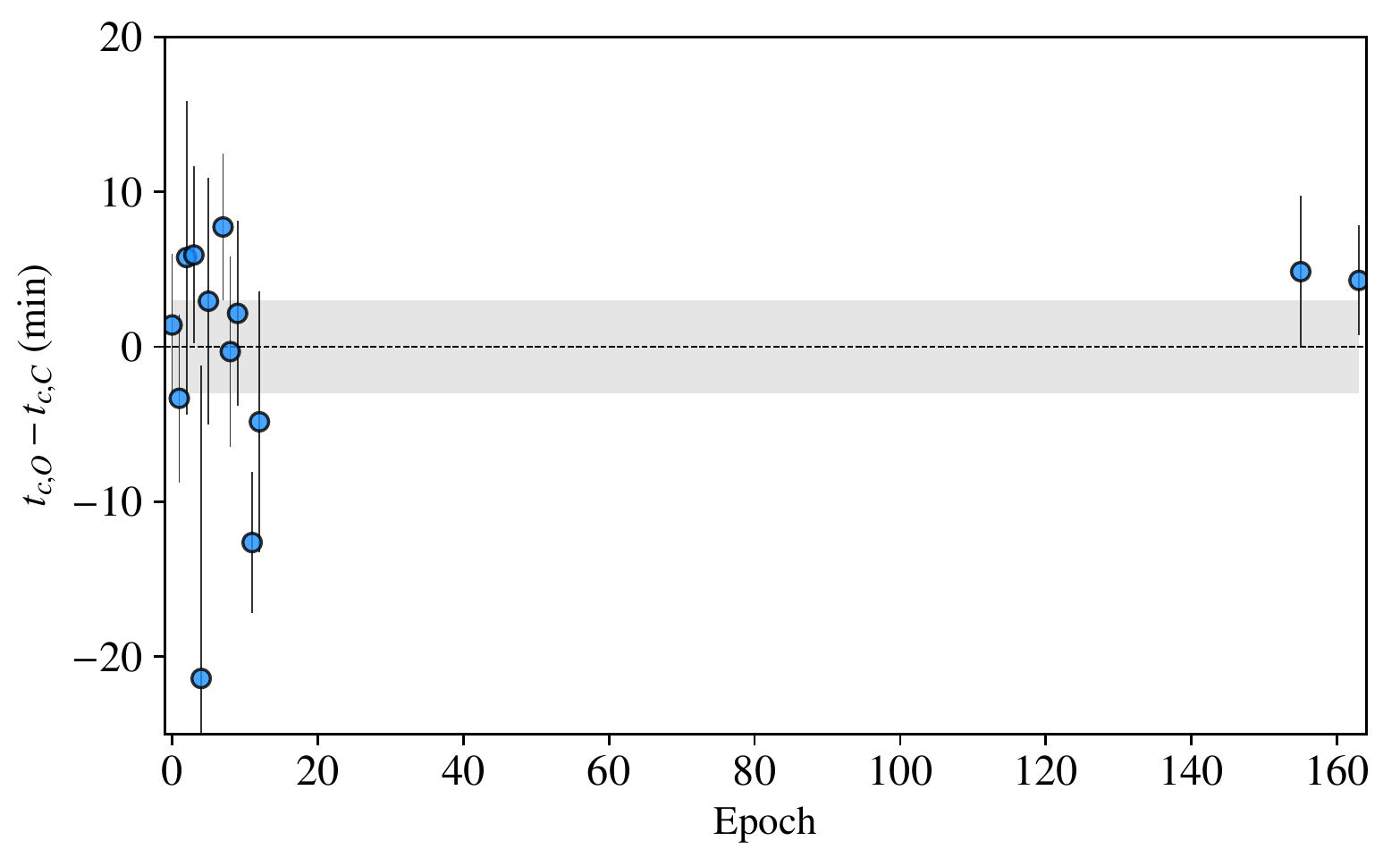}
\caption{Observed minus computed mid-transit times of \planet\,. The residuals (TTV) of the transit times are shown considering the proposed linear ephemeris. The dashed line corresponds to zero variation and the grey area is the propagation of 1$\sigma$ uncertainties, considering the optimal transit time from \texttt{EXOFASTv2}, and the period from \texttt{juliet}. The epoch 0 is the first  transit observed by \tess\, and it is also the corresponding epoch of the optimal transit time. The TTV values shown in this plot fit accordingly with the proposed linear ephemeris ($\chi^2_{red}=1.37$)}
\label{fig:ttvs}
\end{figure}

Considering the theoretical and the observed transit times of the light curves, we obtained the TTV values for TOI-132\,b presented in Figure \ref{fig:ttvs}. Even though the larger variation is about 22 minutes, we found no evidence of a clear periodic variation in the transit time. This outlier is probably induced by a gap in the light curve of epoch 5. The RMS variation from the linear ephemeris is $\sigma = 8.03\ $min, however, the reduced chi-squared for this model is  $\chi^2_{red}=1.37$. This is an indicator that the transit times, considering their errors, fit well with the proposed linear ephemeris. 

The lack of an additional RV signal as well as no evidence of a TTV signal for our given time-span of our transit data, suggest that there is no other close-in companion of TOI-132\,b. Nevertheless, further ground-based follow-up will be required to unveil the possible existence of companions in TOI-132.

\begin{figure}
\centering
\includegraphics[width=\columnwidth]{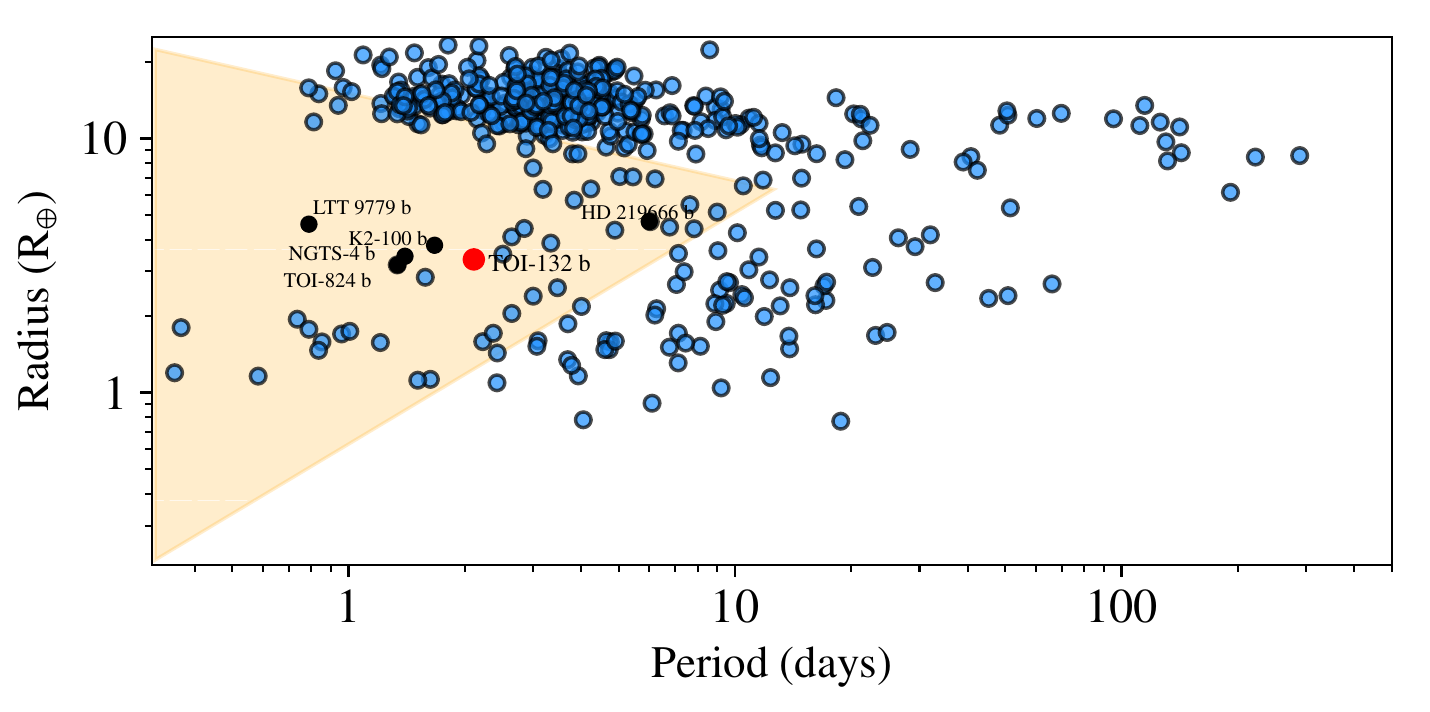}
\includegraphics[width=\columnwidth]{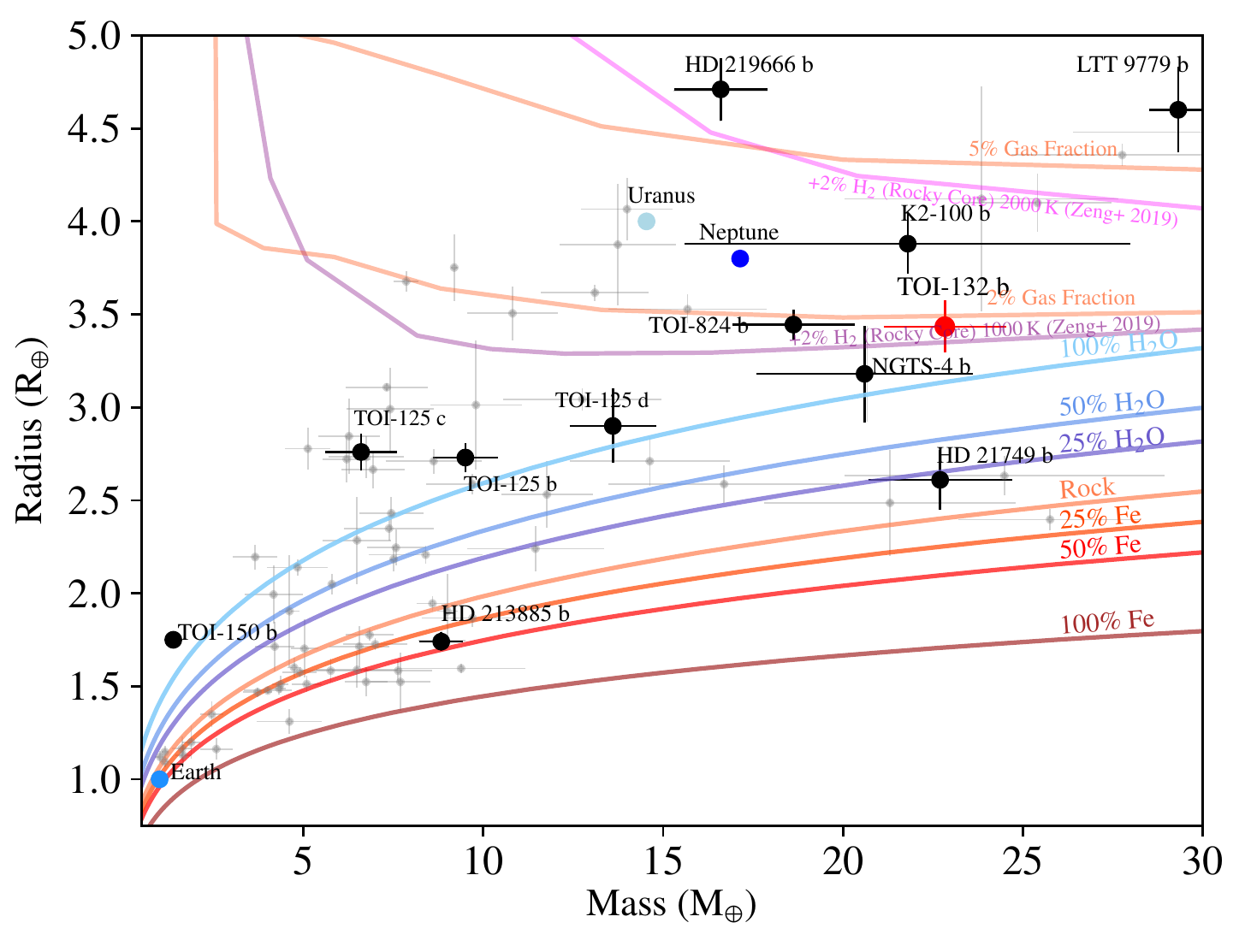}
\caption{{\it Top}: Period-radius diagram for planets whose radius has been measured with a precision better than 5\%. We have included recent {\it TESS} discoveries (Burt et al. 2019, Nielsen et al. 2019, private communication). The shaded area indicates the Neptune-desert where the edges are defined by \citet{Mazeh2016}. TOI-132\,b is highlighted with a red circle, near the edge of the desert. {\it Bottom}: Mass-radius diagram for planets whose mass and radius have been measured  with a precision better than 25\% (gray circles) in the range $R_\mathrm{p} < 5 R_{\oplus}$ and $M_\mathrm{p}< 30 M_{\oplus}$, retrieved from the transiting planets catalog \texttt{TEPCat} (available at \url{https://www.astro.keele.ac.uk/jkt/tepcat/}, \citealp{Southworth2011}). Black points show recent discoveries from {\it TESS}. TOI-132\,b is shown with a red circle. Solid, colored lines show models for different compositions from \citet{Zeng2016} ranging from 100\% iron core planet to 100\% H$_{\rm 2}$O planet. Also two-layer models from \citet{Zeng2019} are shown for 2\% H$_{2}$ envelopes at different temperatures (magenta, purple). Extended models from \citep{LopezFortney2014} are shown for 95\% and 98\% core mass fraction, 6.2 Gyr (orange).} 
\label{fig:mrplot}
\end{figure}

\section{Discussion and Conclusions}

By combining \tess\, space-based photometry with HARPS high-precision radial-velocity measurements, along with additional high-sensitivity ground-based photometric observations, we were able to confirm a short period, hot Neptune-like planet orbiting the nearby metal-rich G8 V star TOI-132. The planet was found to have an orbital period of only 2.1~days, a radius of 3.43$^{+0.13}_{-0.14}$~\re, and mass of 22.83$^{+1.81}_{-1.80}$~\me, implying a density and equilibrium temperature of  3.108$^{+0.44}_{-0.45}$\,\gcm\, 1384$^{+53}_{-75}$~K, respectively.

In Fig.~\ref{fig:mrplot} we can see that TOI-132\,b is located in an underpopulated region of the mass-radius diagram. Of the relatively small number of known Neptune-like planets with well constrained properties, TOI-132\,b stands out as bridging the gap between 100\% water worlds and more typical Neptunes that have atmospheric mass fractions of $\sim$10\%. The planet likely more closely resembles NGTS-4\,b \citep{West2019}, which is shown in the figure despite the relatively high uncertainties measured for the planetary parameters, or TOI-824\,b (Burt et al. 2019, private communication). These three planets appear to have similar masses and radii, giving rise to similar densities and bulk compositions, which might indicate they share similar formation histories. 

Moreover, it is interesting to mention the planet K2-100\,b from the {\it K2} mission \citet{Mann2017}. Recently characterized by \citep{Barragan2019b}, the planet consists of a young, inflated Neptune on a short period around a G-type star. TOI-132\,b falls within the evolutionary range of K2-100\,b after 5 Gyr. This may indicate in the past TOI-132\,b could have shared similar characteristics to that of K2-100\,b, and at some point given the strong stellar irradiation on TOI-132\,b could have caused atmospheric loss we see in the present.

While TOI-132\,b is not as extreme in some respects as the recently discovered, first ultra hot Neptune LTT 9779\,b (Jenkins et al. 2019), it is placed right at the edge of the Neptune desert. The survival of the planet's atmosphere can likely be understood based on its large core mass, and also the incompatibility with being composed of either 100\% rock or water. This would imply that, at the present time, TOI-132\,b could maintain some significant gaseous atmosphere. We employed a 1-D thermal evolution model \citep{LopezFortney2014}, and for an Earth-like rocky core we find a best-fit current day atmospheric mass fraction of 4.3$^{+1.2}_{-2.3}$\% gas, which can be retained with an initial envelope fraction of $\sim9\%$ at 10 Myr.

With the \Gaia\, parameters from Table \ref{tab:star}, we calculated the star's Galactic space motion. We used the \texttt{IDL} routine \texttt{calc\_uvw}, based upon \citet{JohnsonSoderblom1987} and the local standard of rest from \citet{Coskonolu2011}, we obtained (U,V,W)=(18.4 $\pm$ 0.2, -32.6 $\pm$ 0.4, 16.5 $\pm$ 0.4) km s$^{-1}$. Per the methodology of \citet{Reddy2006}, this corresponds to a 98\% probability that TOI-132 belongs to the Galactic thin disk, which is consistent with the relatively high [Fe/H] we measured for the star. 

The relatively high metallicity of the host star can also help to explain the large core mass fraction of the planet.  Such metal-rich disks can quickly build up high-mass cores that can accumulate large fractions of gas before the disk is dispersed on timescales of $\sim$5$\--$10~Myrs \citep{Baraffe2010,Mulders2018}.  Indeed, we may expect more cores to have been formed in this process, possibly influencing the migration history of TOI-132\,b, and therefore future precision radial-velocity measurements should be sought to search for the presence of a more rich planetary system.\\

In this paper, we have presented the {\it TESS} discovery of a Neptune-sized planet transiting the G-type star TOI-132 on the edge of the Neptune desert. Confirmation of this candidate comes from high precision HARPS spectroscopic observations that fully constrain the orbital and physical parameters of TOI-132\,b. Additional ground-based photometry and speckle images provide evidence of the planetary nature of TOI-132\,b. 
Structure models suggest that the planet can have a rocky core, retaining an atmospheric mass fraction of 4.3$^{+1.2}_{-2.3}$\%.  TOI-132\,b stands as a \tess\, Level 1 Science Requirement candidate, which aims to precisely measure the masses for 50 transiting planets smaller than 4$R_{\oplus}$.  Therefore, future follow-up  observations will allow the search for additional planets in the TOI-132 system, and also will help to constrain low-mass planet formation and evolution models, key to understanding the Neptune desert.

\section*{Acknowledgements}
MRD acknowledges support of CONICYT-PFCHA/Doctorado Nacional-21140646 and Proyecto Basal AFB-170002. JSJ is supported by FONDECYT grant 1161218 and partial support from CATA-Basal (PB06, Conicyt). ZMB acknowledges funds from CONICYT-FONDECYT/Chile Postdoctorado 3180405. JIV acknowledges support of CONICYT- PFCHA/Doctorado Nacional-21191829, Chile. M.E. acknowledges the  support  of  the  DFG  priority  program SPP 1992 ``Exploring  the  Diversity  of  Extrasolar  Planets'' (HA 3279/12-1). KWFL, SzCs acknowledge support by DFG grant RA 714/14-1 within the DFG Schwerpunkt SPP 1992, ``Exploring  the  Diversity  of  Extrasolar  Planets''. Funding for the {\it TESS} mission is provided by NASA's Science Mission directorate. NN acknowledges support by JSPS KAKENHI Grant Numbers JP18H01265 and JP18H05439, and JST PRESTO Grant Number JPMJPR1775. AS acknowledges financial support from the French Programme National de Plan\'etologie (PNP, INSU). We acknowledge the use of {\it TESS} Alert data, from pipelines at the {\it TESS} Science Office and at the {\it TESS} Science Processing Operations  Center. Resources supporting this work were provided by the NASA High-End Computing (HEC) Program through the NASA Advanced Supercomputing (NAS) Division at Ames Research Center for the production of the SPOC data products. This research has made use of the Exoplanet Follow-up Observation Program website, which is operated by the California Institute of Technology, under contract with the National Aeronautics and Space Administration under the Exoplanet Exploration Program. This work makes use of observations from the LCO network.

\bibliographystyle{mnras}
\bibliography{references.bib}
\appendix
$^{1}$Departamento de Astronom\'ia, Universidad de Chile, Camino El Observatorio 1515, Las Condes, Santiago, Chile.\\
$^{2}$Dipartimento di Fisica, Universit\`a degli Studi di Torino, via Pietro Giuria 1, I-10125, Torino, Italy.\\
$^{3}$NASA Goddard Space Flight Center, Greenbelt, MD, USA.\\
$^{4}$School of Physics and Astronomy, Queen Mary University of London, G.O. Jones Building, 327 Mile End Road London, E1 4NS, UK.\\
$^{5}$Vanderbilt University, Department of Physics \& Astronomy, 6301 Stevenson Center Ln., Nashville, TN 37235, USA.\\
$^{6}$Fisk University, Department of Physics, 1000 17th Ave. N., Nashville, TN 37208, USA.\\
$^{7}$Center for Astrophysics ${\rm \mid}$ Harvard {\rm \&} Smithsonian, 60 Garden Street, Cambridge, MA 02138, USA.\\
$^{8}$Dunlap Institute for Astronomy and Astrophysics, University of Toronto, Ontario M5S 3H4, Canada.\\
$^{9}$ Department of Space, Earth and Environment, Chalmers University of Technology, Onsala Space Observatory, SE- 439 92 Onsala, Sweden.\\
$^{10}$ Leiden Observatory, University of Leiden, PO Box 9513, 2300 RA, Leiden, The Netherlands.\\
$^{11}$Dept.\ of Physics \& Astronomy, Swarthmore College, Swarthmore PA 19081, USA.\\
$^{12}$ Instituto de Astrof\'isica de Canarias, V\'ia L\'actea s/n, E-38205 La Laguna, Tenerife, Spain.\\
$^{13}$ Departamento de Astrof\'isica, Universidad de La Laguna, Spain.\\
$^{14}$ Aix Marseille Univ, CNRS, CNES, LAM, Marseille, France.\\
$^{15}$ University of Warwick, Department of Physics, Gibbet Hill Road, Coventry, CV4 7AL, UK.\\
$^{16}$ Th{\"u}ringer Landessternwarte  Tautenburg, Sternwarte 5, 07778, Tautenburg, Germany.\\
$^{17}$ Las Cumbres Observatory, 6740 Cortona Dr., Ste. 102, Goleta, CA 93117, USA.\\
$^{18}$ Center for Astronomy and Astrophysics, Technical University Berlin, Hardenbergstr. 36, 10623 Berlin, Germany.\\
$^{19}$ Department of Astronomy, University of Tokyo, 7-3-1 Hongo, Bunkyo-ku, Tokyo 113-0033, Japan.\\
$^{20}$ Mullard Space Science Laboratory, University College London, Holmbury St Mary, Dorking, Surrey RH5 6NT, UK.\\
$^{21}$ Department of Astrophysical Sciences, Princeton University, 4 Ivy Lane, Princeton, NJ 08540, USA.\\
$^{22}$Astrobiology Center, 2-21-1 Osawa, Mitaka, Tokyo 181-8588, Japan.\\
$^{23}$JST, PRESTO, 2-21-1 Osawa, Mitaka, Tokyo 181-8588, Japan.\\
$^{24}$National Astronomical Observatory of Japan, 2-21-1 Osawa, Mitaka, Tokyo 181-8588, Japan.\\
$^{25}$Cerro Tololo Inter-American Observatory, Casilla 603, La Serena, Chile.\\
$^{26}$ George Mason University, 4400 University Drive, Fairfax, VA, 22030, USA.\\
$^{27}$ German Aerospace Center, Institute of Planetary Research, 12489 Berlin, Rutherfordstrasse 2, Germany.\\
$^{28}$MIT Kavli Institute for Astrophysics and Space Research, Massachusetts Institute of Technology, 77 Massachusetts Avenue, 37-241, Cambridge MA 02139, USA.\\
$^{29}$Physics Department and Tsinghua Centre for Astrophysics, Tsinghua University, Beijing 100084, China.\\
$^{30}$ Department of Earth, Atmospheric and Planetary Sciences, Massachusetts Institute of Technology, Cambridge, MA 02139, USA.\\
$^{31}$ NASA Ames Research Center, Moffet Field, CA 94035, USA.\\
$^{32}$Department of Physics and Astronomy, University of North Carolina at Chapel Hill, Chapel Hill, NC 27599-3255, USA.\\
$^{33}$ SETI Institute, 189 Bernardo Ave, Suite 200, Mountain View, CA 94043, USA.\\ 
$^{34}$ Department of Aeronautics and Astronautics, MIT, 77 Massachusetts Avenue, Cambridge, MA 02139, USA.\\
USA.\\
$^{35}$ Perth Exoplanet Survey Telescope, Perth, Western Australia.\\
$^{36}$Department of Astronomy, The University of Texas at Austin, 2515 Speedway, Stop C1400 Austin, Texas 78712-1205, USA.\\
\end{document}